\documentclass[a4paper]{llncs}
\usepackage[utf8]{inputenc}
\title{%
    Pre- and post-quantum Diffie--Hellman
    \\
    from groups, actions, and isogenies
}
\author{Benjamin Smith}
\institute{
    Inria 
    \emph{and}
    Laboratoire d'Informatique de l'École polytechnique (LIX),\\
    Université Paris--Saclay, France\\
    \email{smith@lix.polytechnique.fr}
}

\date{\today}

\usepackage{amsmath,amsfonts,amssymb}
\usepackage{url}
\usepackage{hyperref}
\usepackage{doi}
\usepackage{xcolor}
\usepackage{xspace}
\usepackage[all,cmtip]{xy}
\usepackage[ruled,vlined,linesnumbered]{algorithm2e}
\DontPrintSemicolon

\newcommand{\FF}{\ensuremath{\mathbb{F}}}
\newcommand{\FFbar}{\ensuremath{\overline{\mathbb{F}}}}

\newcommand{\QQ}{\ensuremath{\mathbb{Q}}}
\newcommand{\ZZ}{\ensuremath{\mathbb{Z}}}
\newcommand{\EC}{\ensuremath{\mathcal{E}}}
\newcommand{\XC}{\ensuremath{\mathcal{C}}}
\newcommand{\OO}{\ensuremath{\mathcal{O}}}
\newcommand{\thegroupa}{\ensuremath{\mathcal{G}}\xspace}
\newcommand{\thegroup}{\ensuremath{\mathfrak{G}}\xspace}
\newcommand{\thespace}{\ensuremath{\mathcal{X}}\xspace}
\newcommand{\frakg}{\ensuremath{\mathfrak{g}}\xspace}
\newcommand{\frakh}{\ensuremath{\mathfrak{h}}\xspace}
\newcommand{\fraka}{\ensuremath{\mathfrak{a}}\xspace}
\newcommand{\frakb}{\ensuremath{\mathfrak{b}}\xspace}
\newcommand{\frake}{\ensuremath{\mathfrak{e}}\xspace}
\newcommand{\frakell}{\ensuremath{\mathfrak{l}}\xspace}
\newcommand{\Plus}{\ensuremath{+}\xspace}
\newcommand{\Minus}{\ensuremath{-}\xspace}

\newcommand{\Pic}{\ensuremath{\mathrm{Pic}}}
\newcommand{\End}{\ensuremath{\mathrm{End}}}

\newcommand{\Cl}{\ensuremath{\mathrm{Cl}}}
\newcommand{\dualof}[1]{\ensuremath{{#1}^\dagger}}

\newcommand{\subgrp}[1]{\ensuremath{\langle{#1}\rangle}}

\newcommand{\softO}{\ensuremath{\widetilde{O}}}

\newcommand{\samples}{\ensuremath{\stackrel{\$}{\gets}}\xspace}
\SetKwProg{Fn}{function}{}{}
\SetKwFunction{KeyPair}{KeyPair}
\SetKwFunction{Encode}{Encode}
\SetKwFunction{Decode}{Decode}
\SetKwFunction{DH}{DH}
\SetKwFunction{DHEncap}{DHEncapsulate}
\SetKwFunction{DHDecap}{DHDecapsulate}
\SetKwFunction{ElGamalEnc}{ElGamalEncrypt}
\SetKwFunction{ElGamalDec}{ElGamalDecrypt}
\SetKwFunction{KDF}{KDF}
\SetKwFunction{Enc}{Enc}
\SetKwFunction{Dec}{Dec}
\SetKwFunction{CRT}{CRT}
\SetKwFunction{BSGS}{BSGS}
\SetKwFunction{PohligHellman}{PohligHellman}

\begin{document}
\maketitle

\begin{abstract}
    Diffie--Hellman key exchange 
    is at the foundations of public-key cryptography,
    but conventional group-based Diffie--Hellman
    is vulnerable to Shor's quantum algorithm.
    A range of ``post-quantum Diffie--Hellman''
    protocols have been proposed
    to mitigate this threat,
    including the Couveignes, Rostovtsev--Stolbunov,
    SIDH, and CSIDH schemes,
    all based on the combinatorial and number-theoretic structures 
    formed by isogenies of elliptic curves.
    Pre- and post-quantum Diffie--Hellman schemes
    resemble each other at the highest level,
    but the further down we dive, the more differences
    emerge---differences that are critical when we use Diffie--Hellman
    as a basic component in more complicated constructions.
    In this survey we compare and contrast pre- and post-quantum
    Diffie--Hellman algorithms,
    highlighting some important subtleties.
\end{abstract}

\section{
    Introduction
}
\label{sec:intro}

The Diffie--Hellman key-exchange protocol is,
both literally and figuratively,
at the foundation of public-key cryptography.
The goal is for two parties,
Alice and Bob,
to derive a shared secret from each other's public keys
and their own private keys.
Diffie and Hellman's original solution~\cite{Diffie--Hellman}
is beautifully and brutally simple:
given a fixed prime \(p\) and a primitive element \(g\) in the finite field \(\FF_p\)
(that is, a generator of the multiplicative group \(\FF_p^\times\)),
Alice and Bob choose secret keys \(a\) and \(b\), respectively,
in \(\ZZ/(p-1)\ZZ\).
Alice computes and publishes her public key \(A = g^a\),
and Bob his public key \(B = g^b\);
the shared secret value is \(S = g^{ab}\), 
which Alice computes as \(S = B^a\),
Bob as \(S = A^b\).

The security of the shared secret
depends on the hardness of the 
\emph{Computational Diffie--Hellman Problem} (CDHP),
which is to compute \(S\) given only \(A\), \(B\),
and the public data of the structures that they belong to.
For finite-field Diffie--Hellman,
this means computing \(g^{ab}\) given only \(g\), \(g^a\), and \(g^b
\pmod{p}\).
The principal approach to solving the CDHP
is to solve the \emph{Discrete Logarithm Problem} (DLP),
which is to compute \(x\) from \(g\) and \(g^x\).
We thus recover \(a\) from \(A = g^a\)
(or, equivalently, \(b\) from \(B = g^b\)),
then power \(B\) by \(a\) (or \(A\) by \(b\))
to recover \(S\).
Attacking the DLP means directly attacking one of the public keys,
regardless of any particular shared secret they may be used to derive.

Over the past four decades,
the Diffie--Hellman protocol
has been generalized 
from multiplicative groups of finite fields to a range of other algebraic groups,
most notably elliptic curves.
Partly motivated by this cryptographic application,
there has been great progress in discrete logarithm algorithms
for some groups,
most notably Barbulescu, Gaudry, Joux, and Thomé's
quasipolynomial algorithm for discrete logarithms
in finite fields of fixed tiny characteristic~\cite{BGJT}.

The most stunning development in discrete logarithm algorithms
came with the rise of the quantum computation paradigm:
Shor's quantum algorithm~\cite{Shor94}
solves the discrete logarithm problem---and thus
breaks Diffie--Hellman---in any group 
in polynomial time and space on a quantum computer.\footnote{%
    More generally,
    Armknecht, Gagliardoni, Katzenbeisser, and Peter
    have shown that no group-homomorphic cryptosystem
    can be secure against a quantum adversary,
    essentially because of the existence of Shor's algorithm~\cite{AGKP}.
}
The development of quantum computers of even modest capacity
capable of running Shor's algorithm
remains an epic challenge in experimental physics:
at the time of writing,
the largest implementation of Shor's algorithm
was used to factor the integer 21,
so there is some way to go~\cite{factoring-21}.
But in anticipation,
cryptographic research has already bent itself 
to the construction of \emph{post-quantum} cryptosystems,
designed to be used on conventional computers
while resisting known quantum attacks.

Nowadays, Diffie--Hellman is often an elementary component of a more complicated
protocol, rather than the entire protocol itself.
For example, the TLS protocol used to establish secure internet connections
includes an ephemeral Diffie--Hellman~\cite{TLS13}.
But to give a more interesting example,
the X3DH protocol~\cite{X3DH} used to establish connections
in Signal and WhatsApp
includes \emph{four} simple Diffie--Hellmans
between various short and long-term keypairs.
The common use of Diffie--Hellman as a component
makes the search for a drop-in post-quantum replacement for
classical Diffie--Hellman particularly relevant today.

While many promising post-quantum candidates 
for public-key encryption and signatures
have been developed%
---the first round of the NIST post-quantum standardization process~\cite{NIST-PQ}
saw 59 encryption/KEM schemes and 23 signature schemes submitted---%
finding 
a simple post-quantum drop-in replacement for Diffie--Hellman
(as opposed to a KEM)
has proven to be surprisingly complicated.
Some interesting post-quantum ``noisy Diffie--Hellman'' key exchange protocols
based on hard problems in codes, lattices, and Ring-LWE
have been put forward over
the years
(including~\cite{AGLSZ}, \cite{Ding}, \cite{Ding--Xie--Lin}, \cite{Peikert},
\cite{NewHope}, \cite{BCNS}, and~\cite{Ouroboros}),
but these typically require a reconciliation phase to ensure 
that Alice and Bob 
have the same shared secret value
(as opposed to an approximate shared secret with acceptable noise on each side);
we will not discuss these protocols further here.
Perhaps surprisingly,
given the loudly trumpeted quantum destruction of elliptic curve
cryptography by Shor's algorithm,
the most serious candidates for post-quantum Diffie--Hellman
come from isogeny-based cryptography,
which is founded in the deeper theory of elliptic curves.

The key idea in moving from conventional elliptic-curve cryptography
to isogeny-based cryptography is that 
points on curves are replaced with entire curves, 
and relationships between points (scalars and discrete logarithms)
are replaced with relationships between curves (isogenies).
Isogeny classes have just enough algebraic structure
to define efficient asymmetric cryptosystems,
but not enough to make them vulnerable to Shor's algorithm.

But what should a ``post-quantum Diffie--Hellman'' scheme be,
and how closely should it match classical Diffie--Hellman
functionalities and semantics?
To what extent can the intuition and theoretical lore
built up over decades of classical Diffie--Hellman
carry over to these new protocols?
This survey is an attempt to begin addressing these questions.
The aim is to help cryptographers, mathematicians, and computer scientists
to understand the similarities and differences between 
classical Diffie--Hellman and the new post-quantum protocols.

\paragraph{The plan.}
We begin with a quick survey of classical group-based Diffie--Hellman
in \S\S\ref{sec:DLP}-\ref{sec:DLP-CDHP}.
We discuss modern elliptic-curve Diffie--Hellman
in~\S\ref{sec:modern};
this dispenses with the underlying group on some levels,
and thus forms a pivot for moving towards post-quantum Diffie--Hellman.
We review
Couveignes' \emph{hard homogeneous spaces} framework 
in~\S\ref{sec:PHS} and~\S\ref{sec:HHS},
before introducing HHS cryptosystems in the abstract
in~\S\ref{sec:HHS-crypto};
we go deeper into the underlying hard problems
in~\S\ref{sec:vec-par} and~\S\ref{sec:PH}.
Moving into the concrete,
we recall basic facts about isogenies in~\S\ref{sec:isogeny-background},
before discussing 
commutative isogeny-based key exchange in~\S\ref{sec:isogeny-DH}
and the stranger SIDH scheme in~\S\ref{sec:SIDH}.
Our focus is mostly constructive,
and our discussion of quantum cryptanalysis will be
purely asymptotic, for reasons discussed in~\S\ref{sec:security}.

\paragraph{Limiting scope.}
The basic Diffie--Hellman scheme is completely unauthenticated:
it is obviously vulnerable to a man-in-the-middle attack where Eve
impersonates Bob to Alice, and Alice to Bob.
Alice and Bob must therefore authenticate each other
outside the Diffie--Hellman protocol,
but we do not discuss authentication mechanisms here.
We also ignore the provable-security aspects of these
protocols,
beyond noting that each has been proven session-key secure 
in Canetti and Krawczyk's
adversarial authenticated-links model~\cite{Canetti--Krawczyk} 
(see~\cite[\S5.3]{DFKS} for a proof for commutative isogeny key
exchange,
and~\cite[\S6]{SIDH2} for SIDH).
As we noted above,
we do not discuss noisy Diffie--Hellman schemes here,
mostly for lack of time and space,
but also because these are further from being drop-in replacements
for classical Diffie--Hellman.
Finally, we must pass over \emph{decision} Diffie--Hellman-based
protocols in silence, partly for lack of space,
but mostly because at this early stage it seems hard to say anything
nontrivial about decision Diffie--Hellman in the post-quantum setting.
We do this with some reluctance: as Boneh declared in~\cite{Boneh-DDH},
``the decision Diffie--Hellman assumption is a gold mine''
(at least for theoretical cryptographers).
Revisiting~\cite{Boneh-DDH}
in the post-quantum setting would be highly interesting,
but this is neither the time nor the place for that investigation.

\paragraph{Notation.}
\label{sec:notation}
We will use abelian groups written additively and multiplicatively,
depending on the context.
To minimise confusion,
we adopt these typographical conventions for groups and elements:
\begin{itemize}
    \item
        \(\thegroupa\) always denotes an abelian group written
        \textbf{additively},
        with group operation \((P,Q) \mapsto P+Q\), 
        inverse \(P \mapsto -P\),
        and identity element \(0\).
    \item
        \(\thegroup\) always denotes an abelian group written
        \textbf{multiplicatively},
        with group operation \((\mathfrak{p},\mathfrak{q}) \mapsto
        \mathfrak{pq}\), 
        inverse \(\mathfrak{p} \mapsto \mathfrak{p}^{-1}\),
        and identity element \(1\).
\end{itemize}

\paragraph{Acknowledgements.}
I am grateful to
Luca De Feo, 
Florian Hess,
Jean Kieffer,
and Antonin Leroux
for the many hours they spent
discussing these cryptosystems with me,
and the organisers, chairs, and community of WAIFI 2018.

\section{
    Abstract groups and discrete logarithms
}
\label{sec:DLP}

Diffie and Hellman presented their key exchange in
the multiplicative group of a finite field,
but nothing in their protocol requires the field structure.
We will restate the protocol in the setting of a general
finite abelian group in~\S\ref{sec:pre-quantum-DH};
but first, we recall some basic facts 
about abstract groups and discrete logarithms.

Let \thegroupa be a finite abelian group of order \(N\)
(written additively, following the convention above).
We can assume \thegroupa is cyclic.
For every integer \(m \pmod{N}\) we have an exponentiation
endomorphism \([m]: \thegroupa \to \thegroupa\),
called \emph{scalar multiplication},
defined for non-negative \(m\) by 
\[
    [m]: P \longmapsto \underbrace{P\Plus\cdots\Plus P}_{m \text{ copies}}
\]
and for negative \(m\) by \([m]P = [-m](\Minus P)\).
We can compute \([m]\)
in \(O(\log m)\) \(\thegroupa\)-operations 
using a variety of addition chains;
typically \(m \sim \#\thegroupa = N\).

The fundamental hard algorithmic problem in \thegroupa
is computing the inverse of the scalar multiplication operation:
that is, computing discrete logarithms.

\begin{definition}[DLP]
    The \emph{Discrete Logarithm Problem} in \(\thegroupa\)
    is, given \(P\) and \(Q\) in \(\subgrp{P} \subseteq \thegroupa\),
    compute an \(x\)
    such that \(Q = [x]P\).
\end{definition}

Any DLP instance in any \thegroupa can always be solved
using \(O(\sqrt{N})\) operations in \thegroupa, 
using (for example) 
Shanks' baby-step giant-step algorithm (BSGS),
which also requires \(O(\sqrt{N})\) space~\cite{Shanks};
Pollard's \(\rho\) algorithm
reduces the space requirement to
\(O(1)\)~\cite{Pollard-rho}.
If \(N\) is composite and its (partial) factorization is known,
then we can do better using the Pohlig--Hellman--Silver
algorithm~\cite{Pohlig--Hellman},
which solves the DLP by reducing to the DLP in subgroups of \thegroupa
(see \S\ref{sec:PH} below).

The DLP enjoys random self-reducibility:
if we have an algorithm that solves DLPs for a large fraction \(1/M\) of all
possible inputs, then we can solve DLPs for all possible inputs
after an expected \(M\) random attempts.
Suppose we want to solve an arbitrary DLP instance \(Q = [x]P\).
We choose a random integer \(r\),
try to solve \(Q' = Q + [r]P = [x+r]P\) for \(x+r\),
and if we succeed then we recover \(x = (x + r) - r\).
After \(M\) randomizations,
we expect to find an \(r\) for which \(Q'\) lands in the set of inputs
to which the algorithm applies.

In the pure abstract, 
we consider \(\thegroupa\) as a black-box group:
operations are performed by oracles,
and elements are identified by (essentially random) bitstrings.
This models the absence of useful information that we could derive
from any concrete representation of \thegroupa.
In this setting, Shoup~\cite{Shoup} has proven that the 
complexity of solving the DLP is not merely
in \(O(\sqrt{N})\), but in \(\Theta(\sqrt{N})\).
But in the real world,
we do not have black-box groups;
every group has a specific concrete element representation
and an explicit algorithmic group law.
The difficulty of the DLP then varies 
with the representation,
as we will see
in~\S\ref{sec:DLP-CDHP}.

\section{
    Pre-quantum Diffie--Hellman
}
\label{sec:pre-quantum-DH}

Now let us consider Diffie--Hellman in the abstract.
Let \(\thegroupa\) be a cyclic group,
and fix a public generator \(P\) of \(\thegroupa\).
Public keys are elements of \(\thegroupa\);
private keys are bitstrings,
interpreted as elements of \(\ZZ/N\ZZ\).
Each (public,private)-keypair \((Q=[x]P,x)\)
presents a DLP instance in~\(\thegroupa\).

The Diffie--Hellman protocol takes place in into two logical phases,
which in practice may be separated by a significant period of time.
In the first phase,
the parties generate their keypairs using
Algorithm~\ref{alg:keypair-group} (\KeyPair):
\begin{itemize}
    \item Alice
        generates her keypair as \((A,a) \gets \KeyPair{}\)
        and publishes \(A\);
    \item Bob
        generates his as \((B,b) \gets \KeyPair{}\)
        and publishes \(B\).
\end{itemize}
In the second phase,
they
compute the shared secret \(S\)
with Algorithm~\ref{alg:DH-group} (\DH):
\begin{itemize}
    \item Alice computes \(S \gets \DH(B,a)\);
    \item Bob computes \(S \gets \DH(A,b)\).
\end{itemize}
Alice and Bob have the same value \(S\) because
\(
    S = [a]B = [b]A = [ab]P
    \,.
\)

\begin{algorithm}
    \caption{Keypair generation for textbook Diffie--Hellman in a group \(\thegroupa =
    \subgrp{P}\) of order \(N\).}
    \label{alg:keypair-group}
    \KwIn{\(()\)}
    \KwOut{A pair \((Q,x)\) in \(\thegroupa\times\ZZ/N\ZZ\) such that \(Q = [x]P\)}
    \Fn{\KeyPair{}}{
        \(x \samples \ZZ/N\ZZ\)
        \;
        \(Q \gets [x]P\)
        \;
        \Return{\((Q,x)\)}
        \;
    }
\end{algorithm}

\begin{algorithm}
    \caption{Textbook Diffie--Hellman key exchange in \thegroupa}
    \label{alg:DH-group}
    \KwIn{A public key \(R\) in \(\thegroupa\), and a private key \(x\) in \(\ZZ/N\ZZ\)}
    \KwOut{A shared secret \(S \in \thegroupa\)}
    \Fn{\DH{\(R,x\)}}{
        \(S \gets [x]R\)
        \;
        \Return{\(S\)}
        \tcp*{To be input to a KDF}
    }
\end{algorithm}

The security of the (entire) shared secret depends on the hardness
of the \emph{Computational Diffie--Hellman Problem} (CDHP) in \(\thegroupa\).
\begin{definition}[CDHP]
    \label{def:CDHP}
    The \emph{Computational Diffie--Hellman Problem} in \thegroupa
    is, given \(P\), \(A = [a]P\), and \(B = [b]P\) in \thegroupa,
    to compute \(S = [ab]P\).
\end{definition}

While it is obvious that an algorithm that 
solves the DLP in \thegroupa can solve the CDHP in \thegroupa,
constructing a reduction in the other direction---that is,
efficiently solving DLP instances given access to an oracle solving CDHP instances---is a
much more subtle matter.
It is now generally believed,
following the work of 
den Boer~\cite{den-Boer},
Maurer and Wolf~\cite{Maurer,Maurer--Wolf},
Muzereau, Smart, and Vercauteren~\cite{MSV},
Bentahar~\cite{Bentahar05},
and Boneh and Lipton~\cite{Boneh--Lipton}
that the DLP and CDHP are equivalent 
for the kinds of \thegroupa that cryptographers use in
practice.\footnote{%
    Suppose \(N\) is prime.
    The Maurer reduction for groups of order \(N\)
    requires an auxiliary elliptic curve
    over \(\FF_N\) whose order is \(B\)-smooth---that is,
    such that every prime dividing the order of the auxiliary curve is
    less than \(B\)---for some small \(B\)
    which determines the efficiency of the reduction.
    If we require polynomially small \(B\),
    then we get a polynomial-time reduction;
    but the hypothesis that such curves exist and can be efficiently
    constructed for arbitrary \(N\)
    is extremely strong, if not wildly overoptimistic.
    If \(B\) is simply smaller than \(N^{1/2}\),
    then we get a reduction that is dominated by the cost of 
    an equivalent DLP calculation,
    which is better than nothing;
    it is not so hard to construct such curves (as Bentahar does).
    The middle ground is to accept subexponential \(B\),
    and hence a subexponential reduction,
    as Muzereau, Smart, and Vercauteren do.
    Brown~\cite{Brown} takes a more constructive approach,
    constructing cryptographic elliptic curves
    equipped with a polynomially smooth auxiliary curve.
}
Since solving DLP instances is the only way we know to solve CDHP
instances,
Diffie--Hellman is generally considered to be a member of
the DLP-based family of cryptosystems.

The shared secret \(S\) is \emph{not} suitable for use as a
key for symmetric cryptosystems\footnote{%
    Some protocols do use the shared secret \(S\) as a key,
    most notably the textbook ElGamal encryption presented
    at the start of~\S\ref{sec:DH-protocols}.
};
rather, it should be treated with a Key Derivation Function (KDF)
to produce a proper symmetric key \(K\).
This essentially hashes the secret \(S\),
spreading the entropy of \(S\) uniformly throughout \(K\),
so deriving any information about \(K\)
requires computing the whole of \(S\),
hence solving a CDHP in \(\thegroupa\).
The indistinguishability of \(S\),
and hence its security as a key,
depends on the weaker \emph{Decisional} Diffie--Hellman Problem,
which is beyond the scope of this article.

The lifespan of keypairs is crucial in Diffie--Hellman-based cryptosystems.
In \emph{ephemeral} Diffie--Hellman,
Alice and Bob's keypairs are unique
to each execution of the protocol.
Ephemeral Diffie--Hellman is therefore essentially interactive.
In contrast, \emph{static} Diffie--Hellman
uses long-term keypairs across many sessions.
Alice may obtain Bob's long-term public key
and complete a Diffie--Hellman key exchange with him---and 
start using the shared secret---without
any active involvement on his part.
Static Diffie--Hellman is therefore an important example of 
a \emph{Non-Interactive Key Exchange} (NIKE)
protocol~\cite{Freire--Hofheinz--Kiltz--Paterson}.

Efficient \emph{public-key validation}---that is, 
checking that a public key was honestly generated---is an important,
if often overlooked, requirement for many Diffie--Hellman systems,
particularly those where keys are re-used.
Suppose Alice derives a shared secret key \(K\)
from a Diffie--Hellman exchange with Bob's public key \(B\),
and then uses \(K\) to communicate with Bob.
A malicious Bob might construct an invalid public key \(B\)
in such a way that \(K\) reveals information about Alice's secret key \(a\).
If \((a,A)\) is ephemeral then Bob has learned nothing useful
about~\(a\), since it will never be used again;
but the keypair \((A,a)\) is to be re-used,
as in static Diffie--Hellman, 
then secret information has been leaked,
and
Alice thus becomes vulnerable to active attacks
(see e.g.~\cite{Lim--Lee} for an example).
Public key validation is simple in a finite field:
it usually suffices to check the order of the element.
Antipa, Brown, Menezes, Struik, and Vanstone
describe the process for elliptic-curve public keys~\cite{ABMSV},
and their methods extended to most curve-based algebraic groups without
serious difficulty.
We will see that this is a more serious problem
in post-quantum systems.

\section{
    Encryption and key encapsulation
}
\label{sec:DH-protocols}

The classic ElGamal public-key encryption scheme~\cite{ElGamal}
is closely related to Diffie--Hellman key exchange.
Its key feature is that messages are viewed as elements of the group \thegroupa,
so adding a random-looking element of \thegroupa
to a message in \thegroupa acts as encryption.

Algorithm~\ref{alg:ElGamal-Alice}
lets Alice encrypt a message to Bob.
Alice first generates an ephemeral keypair \((E,e)\),
completes the Diffie--Hellman on her side using Bob's static public key \(B\)
to compute a shared secret \(S\),
which she
uses as a (symmetric) key to encrypt an element of \thegroupa 
via \(M \mapsto C = M + S\)
(with corresponding decryption \(C \mapsto M = C - S\)).
Sending her ephemeral public key \(E\) together with the ciphertext 
allows Bob to compute \(S\) and decrypt with Algorithm~\ref{alg:ElGamal-Bob}.
Since the secret key here
is the bare shared secret \(S\),
untreated by a KDF,
the security of this protocol depends not on the CDHP
but rather on the (easier) decisional Diffie--Hellman problem in~\thegroupa.

\begin{algorithm}
    \caption{Classic ElGamal encryption: Alice encrypts to Bob}
    \label{alg:ElGamal-Alice}
    \KwIn{Bob's public key \(B\) and a message \(M\in \thegroupa\)}
    \KwOut{An element \(E \in \thegroupa\) and a ciphertext \(C\in\thegroupa\)}
    \Fn{\ElGamalEnc{\(B,M\)}}{ 
        \((E,e) \gets \KeyPair{}\)
        \tcp*{\(E = [e]P\)}
        \(S \gets \DH(B,e)\)
        \tcp*{\(B = [b]P \implies S = [eb]P\)}
        \(C \gets M \Plus S\)
        \tcp*{\(C = M \Plus [eb]P\)}
        \Return{\((E,C)\)}
        \;
    }
\end{algorithm}

\begin{algorithm}
    \caption{Classic ElGamal decryption: Bob decrypts from Alice}
    \label{alg:ElGamal-Bob}
    \KwIn{An element \(E\in\thegroupa\), a ciphertext \(M\in\thegroupa\), and Bob's private key \(b\)}
    \KwOut{A plaintext messge \(M\in\thegroupa\)}
    \Fn{\ElGamalDec{\((E,C,b)\)}}{ 
        \(S \gets \DH(E,b)\)
        \tcp*{\(E = [e]P \implies S = [eb]P\)}
        \(M \gets C \Minus S\)
        \tcp*{\(C = M\Plus [eb]P\)}
        \Return{\(M\)}
        \;
    }
\end{algorithm}

It is important to note that this cryptosystem
does not provide semantic security.
For example,
if \((E_1,C_1)\) and \((E_2,C_2)\) 
are encryptions of \(M_1\) and \(M_2\), respectively,
then \((E_1\Plus E_2,C_1\Plus C_2)\) 
is a legitimate encryption of \(M_1\Plus M_2\).
While this property is desirable for certain applications
(such as E-Voting~\cite{Benaloh-simple}),
in most contexts textbook ElGamal cannot be safely used for public-key encryption.

The homomorphic nature of the scheme 
is due to the fact that the group law \Plus
is being used as the encryption and decryption algorithm.
But even if this behaviour is actually desired,
requiring the message to be an element of~\thegroupa
poses two further problems.
First, 
it imposes a hard and inconvenient limit on the size of the message space;
second, it requires an efficient encoding of messages to group elements
(and an efficient decoding to match).
At first glance, 
this second requirement seems straightforward 
for ElGamal instantiated in \(\FF_p^\times\),
since bitstrings of length \(\le \log_2p\) can be immediately
interpreted as integers in \(\ZZ/p\ZZ\),
and hence elements of \(\FF_p\);
but this embedding does not map into the prime-order subgroups where
the protocol typically operates.
This complication is worse in the elliptic curve setting,
where the message-length limit is even more restrictive.

The modern, semantically secure version of ElGamal encryption
is an example of hybrid encryption,
best approached through the more general setting
of Key Encapsulation Mechanisms (KEMs)
and Data Encryption Mechanisms (DEMs)~\cite{Cramer--Shoup,Hofheinz--Kiltz}.
We establish encryption keys using an asymmetric system (the KEM),
before switching to symmetric encryption for data transport (the DEM).

Algorithms~\ref{alg:DH-KEM-Alice}
and~\ref{alg:DH-KEM-Bob}
illustrate a simple Diffie--Hellman-based KEM.
In Algorithm~\ref{alg:DH-KEM-Alice},
Bob has already generated a long-term keypair 
\((B,b)\) and published~\(B\).
Alice takes \(B\),
generates an ephemeral keypair \((E,e)\),
completes the Diffie--Hellman on her side,
and derives a cryptographic key \(K\)
from the shared secret~\(S\).
She can use \(K\) to encrypt messages to Bob\footnote{%
    If Alice immediately encrypts a message under \(K\)
    and sends the ciphertext to Bob with \(E\),
    then this is ``hashed ElGamal'' encryption
    (see~\cite{Abdalla--Bellare--Rogaway} 
    for a full encryption scheme in this style).
},
while \(E\) encapsulates \(K\) for transport.
To decapsulate \(E\) and decrypt messages from Alice,
Bob follows Algorithm~\ref{alg:DH-KEM-Bob},
completing the Diffie--Hellman on his side
and deriving the cryptographic key \(K\)
from the shared secret \(S\).

\begin{algorithm}
    \caption{DH-based KEM: Alice encapsulating to Bob.}
    \label{alg:DH-KEM-Alice}
    \KwIn{Bob's public key \(B\in\thegroupa\)}
    \KwOut{A symmetric encryption key \(K\in\{0,1\}^n\) and an
    encapsulation \(E\in\thegroupa\) of \(K\) under \(B\)}
    \Fn{\DHEncap{\(B\)}}{ 
        \((E,e) \gets \KeyPair{}\)
        \tcp*{\(E = [e]P\)}
        \(S \gets \DH(B,e)\)
        \tcp*{\(S = [eb]P\)}
        \(K \gets \KDF(E\parallel S)\)
        \tcp*{\(K = \KDF(E\parallel[eb]P)\)}
        \Return{\((K,E)\)}
        \;
    }
\end{algorithm}

\begin{algorithm}
    \caption{DH-based KEM: Bob decapsulating from Alice.}
    \label{alg:DH-KEM-Bob}
    \KwIn{An encapsulation \(E\in\thegroupa\) of a symmetric key \(K \in
    \{0,1\}^n\) under Bob's public key \(B\in\thegroupa\),
    and Bob's private key \(b\in\ZZ/N\ZZ\)}
    \KwOut{A symmetric encryption key \(K\in\{0,1\}^n\)}
    \Fn{\DHDecap{\(E,b\)}}{
        \(S \gets \DH(E,b)\)
        \tcp*{\(E = [e]P \implies S = [eb]P\)}
        \(K \gets \KDF(E\parallel S)\)
        \tcp*{\(E = [e]P \implies K = \KDF([e]P\parallel[eb]P)\)}
        \Return{\(K\)}
        \;
    }
\end{algorithm}

\begin{remark}
    While KEMs provide a convenient API and formalism for key establishment,
    they cannot always be used as a replacement for
    plain-old Diffie--Hellman, especially as
    a component in more complicated protocols.
\end{remark}

\section{
    Concrete groups and discrete logarithms
}
\label{sec:DLP-CDHP}

So far, everything has been presented in the abstract;
but if we want to use any of these schemes in practice,
then we need to choose a concrete group \thegroupa.
As we noted in~\S\ref{sec:DLP},
the hardness of the DLP (and hence the CDHP)
varies according to the representation of \thegroupa,
and may fall far short of the \(O(\sqrt{N})\) ideal.
Here we give a very brief overview of the main candidate groups
for group-based Diffie--Hellman,
and DLP-based cryptography in general.
We refer the reader to Guillevic and Morain's excellent
survey~\cite{Guillevic--Morain}
for further detail on discrete logarithm algorithms.

The DLP in prime finite fields, as used by Diffie and Hellman,
is subexponential:
the General Number Field Sieve~\cite{GNFS}
solves DLP instances in \(\FF_p\)
in time \(L_p[1/3,(64/9)^{1/3}]\).\footnote{%
    Recall that
    \(L_X[\alpha,c] = \exp((c + o(1))(\log X)^\alpha(\log\log
    X)^{1-\alpha})\).
}
In extension fields of large characteristic, 
or when the characteristic has a special form,
the complexity is lower,
while still subexponential
(see~\cite{Guillevic--Morain});
in the extreme case of extension fields of tiny characteristic,
the DLP is quasipolynomial in the field size~\cite{BGJT}.
These algorithms can also be used to attack DLPs in algebraic tori,
which are compact representations of subgroups of
\(\FF_q^\times\)
which offer smaller key sizes and efficient arithmetic~\cite{Rubin--Silverberg,XTR}.

Elliptic curves have long been recognised by number theorists as a
generalization of the multiplicative group
(indeed, both the multiplicative and additive groups 
can be seen as degenerate elliptic curves; see e.g.~\cite[\S9]{Cassels}).
Once Diffie and Hellman had proposed their protocol in a multiplicative
group, then, it was perhaps only a matter of time before number theorists
proposed elliptic-curve Diffie--Hellman;
and within a decade Miller~\cite{Miller} and Koblitz~\cite{Koblitz} 
did just this,
independently and almost simultaneously.
The subexponential finite-field DLP algorithms
do not apply to general elliptic curves,
and so far we know of no algorithm with complexity better than
\(O(\sqrt{N})\) for the DLP in a general prime-order elliptic curve.
Indeed,
the only way we know to make use of the geometric structure 
for general curves over prime fields
is to run generic \(O(\sqrt{N})\) algorithms
on equivalence classes modulo \(\pm1\),
but this only improves the running time 
by a factor of roughly \(\sqrt{2}\)~\cite{BLS-negation}.\footnote{%
    More generally, we can work on equivalence classes modulo a
    (sub)group of automorphisms, as in~\cite{Duursma--Gaudry--Morain};
    but in the case of elliptic curves, for any fixed \(\FF_q\),
    there are only two \(\FFbar_q\)-isomorphism classes of curves 
    with automorphisms other than \(\pm1\).
}
We can do better for some elliptic curves defined over some extension
fields~\cite{Gaudry--Hess--Smart,Wiener--Zuccherato,GallantLV00},
and for some small special classes of
curves~\cite{MOV,Balasubramanian--Koblitz,Frey--Ruck,Smart}
(notably pairing-friendly curves);
but in the more than thirty years
since Miller and Koblitz introduced elliptic curve cryptography,
this \(\sqrt{2}\) speedup represents the only real non-quantum algorithmic
improvement for the general elliptic-curve DLP.\footnote{%
    At least, it is the only improvement as far as algorithmic complexity is
    concerned: implementation and distribution 
    have improved substantially.
    It is, nevertheless, quite dumbfounding that in over thirty years of
    cryptographically-motivated research, we have only
    scraped a tiny constant factor away from the classical algorithmic
    complexity
    of the DLP in a generic prime-order elliptic curve over a prime finite field.
}

Going beyond elliptic curves,
a range of other algebraic groups have been proposed for use in
cryptography.
Koblitz proposed cryptosystems in Jacobians of hyperelliptic curves
as an obvious generalization of elliptic curves~\cite{Koblitz-hyperelliptic}.
Others have since suggested Jacobians of general algebraic curves,
and abelian varieties~\cite{Murty,Robert};
but as the genus of the curve (or the dimension of the abelian variety) grows,
index-calculus algorithms become more effective,
quickly outperforming generic DLP algorithms.
At best, the DLP for curves of fixed genus \(\ge 3\) is exponential,
but easier than \(O(\sqrt{N})\)~\cite{GTTD,Smith-g3,Diem--Thome,Gaudry09};
at worst, as the genus and field size both tend to infinity, 
the DLP becomes subexponential~\cite{Enge--Thome}.
Déchène proposed generalized Jacobians as a bridge between 
elliptic-curve and finite-field cryptography~\cite{Dechene},
but these offer no constructive advantage~\cite{Galbraith--Smith}.

The groups mentioned above
are all \emph{algebraic groups}:
elements are represented by tuples of field elements,
and group operations are computed using polynomial formul\ae{}.
Algebraic groups are well-suited to efficient computation
on real-world computer architectures,
but they are not the only such groups:
another kind
consists of class groups of number fields.
Buchmann and Williams proposed Diffie--Hellman schemes 
in class groups of quadratic imaginary orders~\cite{Buchmann--Williams},
leading to a series of DLP-based cryptosystems set in more general rings
(see~\cite{Buchmann--Takagi--Vollmer} for a survey);
but ultimately these are all vulnerable to subexponential
index-calculus attacks. 

In the classical world, therefore,
elliptic curves over \(\FF_p\) and \(\FF_{p^2}\)
and genus-2 Jacobians over \(\FF_{p^2}\)
present the hardest known DLP instances
with respect to group size (and hence key size).
Elliptic curves over prime fields have become, in a sense,
the gold standard to which all other groups are compared.

\section{
    Concrete hardness and security levels
}
\label{sec:security}

It is important to note that our understanding of DLP hardness is not 
just a matter of theory and sterile asymptotics;
all of the algorithms above are backed by a tradition of experimental work.
Recently discrete logarithms have been computed 
in \(768\)-bit general
and \(1024\)-bit special prime fields~\cite{KleinjungDLPS17,FriedGHT17},
and
in \(112\)-bit
and \(117\)-bit binary elliptic curve groups~\cite{WengerW16,BELNPSZ}.
A table of various record finite field discrete logarithm computations
can be found at~\cite{dldb}.

These computations give us confidence in defining cryptographic parameters
targeting concrete security levels against classical adversaries.
For example, it is generally accepted that DLP-based cryptosystems
in \(\FF_p^\times\) with \(\log_2p\approx3072\)
or in \(\EC(\FF_p)\) with \(\log_2p \approx 256\) for a well-chosen \(\EC\)
should meet a classical approximate 128-bit security level:
that is, a classical adversary equipped with current algorithms
should spend around \(2^{128}\) computational resources to break the system with
non-negligable probability.

For quantum adversaries,
we know that DLPs can be solved in polynomial time---but we still know
relatively little about the concrete difficulty and cost of mounting quantum
attacks against DLP-based cryptosystems, let alone against candidate
post-quantum systems.
For example, we mentioned above that 
the current record for Shor's factoring algorithm is 21; 
but to our knowledge, Shor's algorithm for discrete logarithms
has never been implemented.
Roetteler, Naehrig, Svore and Lauter
have estimated the quantum resources required
to compute elliptic-curve discrete logarithms~\cite{RoettelerNSL17},
which is an important first step.

The situation becomes even murkier 
for the quantum algorithms we will meet below,
including the Kuperberg, Regev, Tani, and Childs--Jao--Soukharev
algorithms.
We have asymptotic estimates,
but no concrete estimates (or real data points, for that matter).
It is not clear what the most useful performance metrics are for these algorithms, 
or how to combine those metrics with classical ones
to estimate overall problem difficulty.

For this reason, we will refrain from giving any concrete security
estimates or recommendations for key-lengths for the post-quantum systems
in the second half of this article.
We look forward to detailed theoretical estimates
along the lines of~\cite{RoettelerNSL17},
and to the eventual development of quantum computers
sufficiently large 
to implement these algorithms and get some actual cryptanalysis done.

\section{
    Modern elliptic-curve Diffie--Hellman
}
\label{sec:modern}

At first glance,
elliptic-curve cryptography is just finite-field cryptography 
with a different algebraic group seamlessly swapped in\footnote{%
    Not entirely seamlessly:
    some operations, like hashing into \thegroupa,
    become slightly more complicated 
    when we pass from finite fields to elliptic curves
    (see~\cite{OJRHT}).
},
and no theoretical modification.
But Miller's original article~\cite{Miller}
ends with an interesting observation that departs from
the multiplicative group perspective:
\begin{quotation}
    Finally, it should be remarked,
    that even though we have phrased everything in terms of points on an
    elliptic curve, that, for the key exchange protocol (and other uses
    as one-way functions), that only the \(x\)-coordinate needs to be
    transmitted.
    The formulas for multiples of a point cited in the first section
    make it clear that the \(x\)-coordinate of a multiple depends only
    on the \(x\)-coordinate of the original point.
\end{quotation}
Miller is talking about elliptic curves in Weierstrass models 
\(y^2 = x^3 + ax + b\),
where \(-(x,y) = (x,-y)\),
so \(x\)-coordinates correspond to group elements modulo sign.
The mapping \((m,x(P))\mapsto x([m]P)\) is mathematically well-defined,
because every \([m]\) commutes with \([-1]\);
but it can also be computed efficiently.

In Diffie--Hellman, then,
instead of using
\begin{align*}
    A 
    & = [a]P
    \,,
    &
    B 
    & = [b]P
    \,,
    &
    S 
    & = [ab]P
    \,,
    \intertext{Miller is proposing that we use}
    A 
    & = \pm[a]P
    &
    B 
    & = \pm[b]P
    &
    S 
    & = \pm[ab]P
    \\
    & = x([a]P)
    \,,
    &
    & = x([b]P)
    \,,
    &
    & = x([ab]P)
    \,.
\end{align*}
Clearly, we lose nothing in terms of security by doing this:
the \(x\)-coordinate CDHP reduces immmediately to the CDHP
in the elliptic curve.
Given a CDHP oracle for \(\EC\),
we can compute \(\pm [ab]P\) 
from \((\pm P,\pm[a]P,\pm[b]P)\)
by choosing arbitrary lifts to 
signed points on \(\EC\) and calling the oracle there;
conversely,
given an \(x\)-coordinate CDHP oracle,
we can solve CDHP instances on \(\EC\) by projecting to the \(x\)-line,
calling the oracle there, and then guessing the sign on \(S\).

The idea of transmitting only the \(x\)-coordinates
may seem advantageous in terms of reducing bandwidth,
but in reality, where elliptic curve point keys
are systematically compressed to an \(x\)-coordinate
plus a single bit to indicate the ``sign'' of the corresponding \(y\)-coordinate,
there is little to be gained here beyond avoiding the small effort
required for compression and decompression.
The real practical value in Miller's idea is that
working with only \(x\)-coordinates is faster,
and requires less memory:
\(x([a]P)\) can be computed from \(a\) and \(x(P)\)
using fewer field operations than would be required
to compute \([a]P\) from \(a\) and \(P\).

This advantage was convincingly demonstrated by 
Bernstein's \texttt{Curve25519} software~\cite{Curve25519},
which put Miller's idea into practice
using carefully selected curve parameters
optimized for Montgomery's ladder algorithm,
which computes the pseudo-scalar multiplications
\(x(P) \mapsto x([m]P)\)
using a particularly efficient and regular differential addition
chain~\cite{Montgomery,Costello--Smith}.
The result was not only a clear speed record for Diffie--Hellman at the
128-bit security level,
but a new benchmark in design for key exchange software.
\texttt{Curve25519} is now the Diffie--Hellman to which all others are
compared in practice.

The elliptic curve cryptosystems that were standardized 
in the 1990s and early 2000s, 
such as the so-called NIST~\cite{NIST}
and Brainpool~\cite{Brainpool} curves,
are based on full elliptic curve arithmetic
and are not optimized for \(x\)-only arithmetic.
More recently, \texttt{Curve25519} and similar systems have
been standardized for future internet applications~\cite{RFC7748}.
These systems are also preferred in newer practical applications, 
such as the Double Ratchet algorithm used for key management 
within the Signal protocol for end-to-end encrypted
messaging~\cite{DoubleRatchet}.

In theory, Miller's idea of working modulo signs
(or, more generally, automorphisms)
extends to any algebraic group.\footnote{%
    While quotienting by \(\pm1\)
    is useful in curve-based cryptosystems,
    it is counterproductive in multiplicative groups of finite fields.
    There, the pseudo-scalar multiplication is \((m,P+1/P) \mapsto P^m + 1/P^m\);
    computing this is slightly slower than computing simple exponentiations,
    and saves no space at any point.
}
For example,
the quotients of Jacobians of genus-2 curves
by \(\pm1\) are Kummer surfaces.
Under suitable parametrizations,
these have highly efficient
pseudo-scalar multiplications~\cite{Gaudry-theta},
which have been used in high-speed Diffie--Hellman 
implementations~\cite{BCLS,muKummer}.

While the \(x\)-only approach to elliptic curve Diffie--Hellman 
is particularly useful in practice,
it also highlights an important theoretical point:
on a formal level, 
Diffie--Hellman does \emph{not} require a group structure.\footnote{%
    Buchmann, Scheidler, and Williams later proposed 
    what they claimed was the first group-less key exchange
    in the infrastructure of real quadratic fields~\cite{BSW}.
    Mireles Morales investigated the
    infrastructure in the analogous even-degree hyperelliptic function
    field case~\cite{MM-inf},
    relating it to a subset of the class group of the field; 
    in view of his work, it is more appropriate to describe
    infrastructure key exchange as group-based.
    In any case, coming nearly a decade after Miller, 
    this would not have been the first non-group Diffie--Hellman.
}
By this, we mean that the group law never explicitly appears in the
protocol---and this is precisely why Diffie--Hellman works on
elliptic \(x\)-coordinates, 
where there is no group law 
(but where there \emph{are} maps induced by scalar multiplication).

The group structure is still lurking behind the scenes, of course.
It plays several important roles:
\begin{enumerate}
    \item
        \emph{Correctness}.
        The group law gives an easy proof that the 
        pseudo-scalar multiplication operations \((m,x(P))\mapsto x([m]P)\)
        exist and commute.
    \item
        \emph{Efficiency}.
        The group law induces biquadratic relations on \(x\)-coordinates
        that we use to efficiently compute pseudo-scalar multiplications
        using suitable differential addition chains~\cite{DJB-chain}.
    \item
        \emph{Security}.
        The hardness of the CDHP in the full group
        underwrites the hardness of the \(x\)-coordinate CDHP.
\end{enumerate}

\begin{remark}
    Can we do without a group entirely?
    Heading into the pure abstract,
    we can consider a Diffie--Hellman protocol with minimal algebraic
    structure.
    For example,
    we could take a set \thespace of public keys in place of the group \thegroupa,
    and sample private keys from a set \(\mathcal{F}\)
    of functions \(\thespace\to\thespace\) defined by the property\footnote{%
        If we require this property to hold 
        for \emph{all} \(P\) in \thespace,
        then \(\mathcal{F}\) is a \emph{commutative magma}.
        Diffie--Hellman protocols where
        \(\mathcal{F}\) is equipped with a semigroup or semiring structure
        have been investigated~\cite{MMR},
        though the results are only of theoretical interest.
    }
    \[
        (a\circ b)(P) = (b\circ a)(P) 
        \quad
        \text{for all}
        \quad
        a, b \in \mathcal{F}
        \,.
    \]
    The associated Diffie--Hellman protocol is then
    defined by
    \begin{align*}
        A & = a(P) 
        \,,
        &
        B & = b(P)
        \,,
        &
        S & = a(b(P)) = b(a(P))
        \,.
    \end{align*}
    We need \(\mathcal{F}\) to be large enough 
    to prevent brute force attacks on \(S\);
    we must be able to efficiently sample functions from \(\mathcal{F}\),
    and evaluate them at elements of~\thespace;
    and we need to justify the hardness of the associated CDHP.
    An algebraic structure on \(\mathcal{F}\)
    may not be strictly necessary to ensure all of this,
    but it certainly makes life easier.
\end{remark}

\section{
    Principal homogeneous spaces
}
\label{sec:PHS}

At the time of writing,
the closest thing we have to a post-quantum analogue of Diffie--Hellman
comes from \emph{isogeny-based cryptography},
whose origins go back to
Couveignes' ``Hard Homogeneous Spaces'' manuscript~\cite{Couveignes}.
This went unpublished for ten years, before appearing
online more or less at the same time as
its ideas were independently rediscovered by Rostovtsev and
Stolbunov~\cite{Rostovtsev--Stolbunov}.

Couveignes' work is a convenient framework
for reasoning about isogeny-based cryptosystems:
the hard detail on class groups and isogeny classes
is abstracted away into groups acting on sets.
We warn the reader that from now on we will mostly be
working with abelian groups written \emph{multiplicatively},
which we denote by \thegroup
in accordance with the convention from~\S\ref{sec:notation}.

Recall that a (left) \emph{action} of a group \thegroup
on a set \thespace
is a mapping \(\thegroup\times\thespace\to\thespace\),
written \((\frakg,P)\mapsto\frakg\cdot P\),
compatible with the group operation in \(\thegroup\):
that is,
\[
    \frakg_1\cdot(\frakg_2\cdot P) = (\frakg_1\frakg_2)\cdot P
    \quad
    \text{for all } 
    \frakg_1, \frakg_2 \in \thegroup
    \text{ and }
    P \in \thespace
    \,.
\]
In our case \thegroup is abelian,
so
\(
    \frakg_1\cdot(\frakg_2\cdot P) 
    = \frakg_2\cdot(\frakg_1\cdot P)
\)
for all \(\frakg_1\), \(\frakg_2\), and \(P\).

\begin{definition}[PHS]
    A \emph{principal homogeneous space (PHS)} for an abelian group \thegroup 
    is a set \thespace equipped with a simple, transitive action of \thegroup:
    that is,
    for any \(P\) and \(Q\) in \thespace,
    there is a \emph{unique} \frakg in \thegroup
    such that \(Q = \frakg\cdot P\).
    Equivalently, 
    for every \(P\) in \thespace,
    the map \(\varphi_P: \thegroup \to \thespace\)
    defined by \(\frakg\mapsto\frakg\cdot P\)
    is a bijection.
\end{definition}

\begin{example}
    \label{eg:group-itself}
    The trivial example of a PHS is a group acting on itself via
    its own group operation:
    that is, \(\thespace = \thegroup\),
    with \(\frakg\cdot\fraka = \frakg\fraka\).
\end{example}
    
\begin{example}
    \label{eg:affine}
    The classic first example of a nontrivial PHS is 
    a vector space acting by translation on its underlying affine space.
\end{example}

Example~\ref{eg:affine}
illustrates a classic informal definition:
a PHS is a group whose identity element has been forgotten
(mislaid, not omitted).
Affine spaces have no distinguished ``origin'';
on the other hand, as soon as one is (arbitrarily) chosen,
then each point defines an implicit displacement vector,
and we can identify the affine space with a vector space.
More generally,
for each \(P\) in \(\thespace\),
if we define 
\(
    \varphi_P(\frakg_1)\varphi_P(\frakg_2) 
    := 
    \varphi_P(\frakg_1\frakg_2)
\)
then we get a well-defined group structure on \thespace;
in fact, we get a different group structure for each choice of \(P\).
The idea therefore is not so much that the identity element has been
forgotten, yet might still be remembered;
it is rather that every single
element is an equally plausible identity.

\begin{example}
    \label{eg:torsor}
    Let \thespace be the set of points on a curve \(\XC\) of genus 1,
    and let \(\thegroup =  \Pic^0(\XC)\) 
    be the group of degree-0 divisor classes on \(\XC\).
    By the Riemann--Roch theorem,
    for every class \([D]\) in \(\Pic^0(\XC)\)
    and point \(P\) on \(\XC\),
    there exists a unique 
    \(P_D\) on \(\XC\)
    such that \([D] = [P_D-P]\).
    We therefore have
    an explicit action of \thegroup on \thespace,
    defined by \([D]\cdot P = P_D\).
    If we fix a choice of distinguished ``base point'' \(O\) in \thespace,
    then we can identify each class \([D]\)
    with the point \([D]\cdot O\),
    and thus, by transport of structure,
    we get a group law on~\thespace.
    (Cognoscenti will recognise the definition of the group law on an
    arbitrary elliptic curve via the Picard group.)
\end{example}

Our final example of a PHS is 
fundamental in isogeny-based cryptography.
It is far more complicated to define
than Examples~\ref{eg:group-itself},
\ref{eg:affine},
and~\ref{eg:torsor};
we will give an extremely brief description here,
returning to it in greater detail 
in~\S\ref{sec:isogeny-background} and~\S\ref{sec:isogeny-DH}.

\begin{example}
    \label{eg:isogeny-HHS}
    Let \(q\) be a prime power
    and \(t\) an integer with \(|t|\le2\sqrt{q}\);
    let \(\OO_K\) be the ring of integers
    of the imaginary quadratic field \(K = \QQ(\sqrt{\Delta})\),
    where \(\Delta := t^2 - 4q\).
    Let \thespace be the set of 
    \(\FF_q\)-isomorphism classes of elliptic curves \(\EC/\FF_q\)
    whose \(\FF_q\)-endomorphism ring is isomorphic to \(\OO_K\)
    (and where the image of the Frobenius endomorphism of \(\EC\)
    in \(\OO_K\) has trace \(t\)).
    Then \thespace is a PHS under
    the ideal class group \(\thegroup = \Cl(\OO_K)\) of \(\OO_K\),
    with ideals acting by \(\fraka\cdot\EC = \EC/\EC[\fraka]\),
    where \(\EC[\fraka]\) is the intersection of the kernels
    of the endomorphisms in \(\fraka\).
    This PHS is 
    central to the theory of Complex Multiplication;
    there is also a well-developed algorithmic theory for it,
    used to compute fundamental number-theoretic objects
    such as modular and Hilbert class polynomials 
    (see e.g.~\cite{BrokerLS12}).
\end{example}

Example~\ref{eg:isogeny-HHS}
highlights another view of PHSes:
we can consider \thespace as a version of
\thegroup whose structure is hidden by the maps \(\varphi_P\).
In this case,
the elements of \thespace are \(j\)-invariants,
and (when the class group is sufficiently large) look like random elements of \(\FF_q\);
the class group itself has no such encoding.

\section{
    Hard homogeneous spaces
}
\label{sec:HHS}

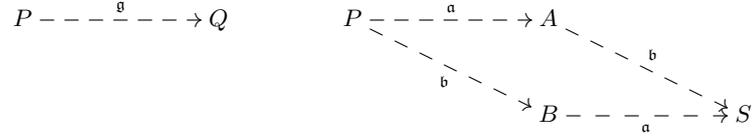
\begin{figure}
    \[
        \xymatrixcolsep{5pc}
        \xymatrix{
            P
            \ar@{-->}[r]^{\mathfrak{g}}
            & 
            Q
        }
        \qquad
        \qquad
        \xymatrix{
            P
            \ar@{-->}[r]^{\mathfrak{a}} 
            \ar@{-->}[rd]_{\mathfrak{b}} 
            & 
            A
            \ar@{-->}[rd]^{\mathfrak{b}}
            \\
            &
            B
            \ar@{-->}[r]_{\mathfrak{a}} 
            & 
            S
        }
    \]
    \caption{Vectorization (left: finding the unique \frakg such that \(Q =
    \frakg\cdot P\)) and parallelization (right: computing the unique \(S\)
    such that \(S = \frakb\cdot A = \fraka\cdot B = (\fraka\frakb)\cdot P\)).
    The dashed arrows denote actions of the unknown group elements
    \(\frakg\), \(\fraka\), and \(\frakb\).}
    \label{fig:vectorization-parallelization}
\end{figure}

Let \thespace be a PHS under \thegroup.
From now on we assume we can efficiently compute group operations,
evaluate actions, test equality, and hash elements
of \thegroup and \thespace.
We also assume we can uniformly randomly sample elements of \thegroup.
Figure~\ref{fig:vectorization-parallelization}
illustrates
the two interesting computational problems in this setting,
which Couveignes called
\emph{vectorization} (Definition~\ref{def:vectorization})
and \emph{parallelization} (Definition~\ref{def:parallelization}).

\begin{definition}[Vectorization]
    \label{def:vectorization}
    The \emph{vectorization} problem in a PHS \thespace under \thegroup is,
    given \(P\) and \(Q\) in \thespace,
    to compute the unique \frakg in \thegroup
    such that \(Q = \frakg\cdot P\).
\end{definition}

\begin{definition}[Parellelization]
    \label{def:parallelization}
    The \emph{parallelization} problem in a PHS \thespace under \thegroup
    is,
    given \(P\), \(A\), and \(B\) in \thespace,
    to compute the unique \(S\) in \thespace
    such that \(S= (\fraka\frakb)\cdot P\)
    where \(A = \fraka\cdot P\)
    and \(B = \frakb\cdot P\).
    (Note that then \(S = \fraka\cdot B = \frakb\cdot A\).)%
\end{definition}
\begin{definition}[HHS]
    Let \thespace be a PHS under \thegroup.
    We say \thespace is a \emph{hard homogeneous space (HHS)}
    if the action of \thegroup on \thespace is efficiently computable, 
    but the vectorization and parallelization problems are
    computationally infeasible.
\end{definition}

The names ``vectorization'' and ``parallelization'' are intuitive 
in the context of Example~\ref{eg:affine}:
vectorization is computing the displacement vector 
between points \(P\) and \(Q\) in the space \thespace, 
while parallelization is completing the parallelogram
with vertices \(P\), \(A\), and \(B\).
These are routine operations in vector and affine spaces,
so the PHS of Example~\ref{eg:affine} is typically not something that we
would consider an HHS.
Similarly, the PHS of Example~\ref{eg:torsor} is not an HHS, 
because we can always vectorize by formally subtracting points to form a degree-0 divisor.
Couveignes suggested that the PHS of Example~\ref{eg:isogeny-HHS}
might be an HHS---and with the current state of classical and quantum
class group and isogeny algorithms, it is.

\section{
    Cryptography in hard homogeneous spaces
}
\label{sec:HHS-crypto}

On a purely symbolic level,
the vectorization problem 
\((P,\frakg\cdot P)\mapsto \frakg\) in a PHS
bears an obvious formal resemblance to the DLP 
\((P,[x]P)\mapsto x\) in a group,
just as the parallelization problem 
\((P,\fraka\cdot P,\frakb\cdot P)\mapsto \fraka\frakb\cdot P\)
resembles the CDHP \((P,[a]P,[b]P)\mapsto [ab]P\).
The presence of abelian groups in each problem
suggests deeper connections---connections that do not necessarily exist.
Indeed, we saw above that parallelization in a PHS is an implicit computation of
the group law, while the Diffie--Hellman operation in a group is something
completely different.

But irrespective of the relationship
between parallelizations and CDHPs,
this syntactical resemblance allows us
to define a cryptosystem analogous to Diffie--Hellman.
Algorithms~\ref{alg:HHS-KeyGen} and~\ref{alg:HHS-DH}
define Couveignes' key exchange in the HHS setting,
with security depending on the hardness of parallelization.

\begin{algorithm}
    \caption{Key generation for cryptosystems in an HHS \thespace under~\thegroup, 
    with a fixed base point \(P\) in \thespace}
    \label{alg:HHS-KeyGen}
    \KwIn{()}
    \KwOut{A private-public keypair $(Q,\frakg)\in
    \thegroup\times\thespace$
    s.t. $Q = \frakg\cdot P$}
    \Fn{\KeyPair{}}{
        $\frakg \gets \texttt{Random}(\thegroup)$
        \tcp*{$\frakg$ is sampled uniformly at random from $\thegroup$}
        $Q \gets \frakg\cdot P$
        \;
        \Return{$(Q,\frakg)$}
        \;
    }
\end{algorithm}

\begin{algorithm}
    \caption{Diffie--Hellman in an HHS \thespace under a group \thegroup}
    \label{alg:HHS-DH}
    \KwIn{A private key $\frakg_A\in \thegroup$ and a public key $Q_B\in
    \thespace$,
    each generated by calls to Algorithm~\ref{alg:HHS-KeyGen}}
    \KwOut{A shared secret value $S\in \thespace$}
    \Fn{\DH{\(Q_B,\frakg_A\)}}{
        $S \gets \frakg_A\cdot Q_B$
        \;
        \Return{$S$}
        \;
    }
\end{algorithm}

Algorithms~\ref{alg:HHS-KeyGen} and~\ref{alg:HHS-DH}
immediately raise some important restrictions on 
the kinds of~\thegroup and~\thespace that we can use.
The first is that we need some kind of canonical representation for
elements of \thespace, to ensure that Alice and Bob can derive equal
shared cryptographic keys from the shared secret \(S\).
This property is also important in settings
where public keys are required to be unique for a given private key.
We also need to be able to efficiently draw
uniformly random samples from~\thegroup;
and then,
given a random element of~\thegroup,
we need to be able to efficiently compute its action on arbitrary
elements of~\thespace.
An alternative approach is to repeatedly randomly sample from a subset
of efficiently-computable elements of \thegroup,
building a sequence of such elements to be used as the secret key,
with the action of the key being the composition of the action of its
components.
This approach requires an argument that the distribution of these compositions
is sufficiently close to the uniform distribution on the whole of \thegroup.
Both approaches are relevant in isogeny-based cryptography,
as we will see in~\S\ref{sec:isogeny-DH}.

Many CDHP-based cryptosystems have obvious HHS analogues.
We can define an HHS-based KEM 
(and implicitly, a hashed-ElGamal-type public key encryption scheme) 
along the lines of 
Algorithms~\ref{alg:DH-KEM-Alice}
and~\ref{alg:DH-KEM-Bob},
by replacing the calls to Algorithms~\ref{alg:keypair-group}
and~\ref{alg:DH-group}
with calls to Algorithms~\ref{alg:HHS-KeyGen}
and~\ref{alg:HHS-DH},
respectively.
But not all DLP-based protocols
have HHS-based analogues:
for example,
the obvious HHS analogue of Schnorr's signature scheme~\cite{Schnorr}
would appear to require an efficient (decisional) parallelization algorithm
in order to verify signatures.

HHS-Diffie--Hellman is \emph{not} 
a natural generalization of group-Diffie--Hellman.
As we noted in~\S\ref{sec:modern},
in group-DH, we have a ring (integers modulo~\(N\))
acting on the group \(\thegroupa\);
the composition operation at the heart of DH is ring multiplication,
but the ring only forms a group under addition.
Formally, in group-DH
we only use the fact that the scalars form a commutative magma;
but algorithmically,
we exploit the fact that the elements form an abelian group 
and the scalars form a commutative ring,
mapping addition in the ring onto the group law,
in order to efficiently evaluate scalar multiplications
using addition chains.

More concretely, 
we noted in~\S\ref{sec:PHS} that the maps 
\(\varphi_P: \frakg\mapsto\frakg\cdot P\)
can be seen as hiding the group \thegroup in \thespace.
The parallelization 
\((P,\fraka\cdot P, \frakb\cdot P) \mapsto \fraka\frakb\cdot P\)
can be written as
\((\varphi_P(1),\varphi_P(\fraka),\varphi_P(\frakb))\mapsto\varphi_P(\fraka\frakb)\);
that is, parallelization 
computes the group law in the hidden representation of
\thegroup in \thespace corresponding to \(P\).
From this perspective, HHS-Diffie--Hellman is a hidden version 
of the ridiculous key exchange where the shared secret is the product of
the two public keys: obviously, without the hiding, this offers no
security whatsoever.

\section{
    Vectorization and parallelization
}
\label{sec:vec-par}

To argue about the security of the schemes in~\S\ref{sec:HHS-crypto},
we must address the following questions: 
how hard are vectorization and parallelization?  
What is the relationship between these problems,
and to what extent does our common intuition 
relating the DLP and CDHP carry over to vectorization and
parallelization in the PHS setting?

It might seem excessive to require the simple and
transitive action of a PHS in all this: we could relax and set up the
same cryptosystems with a group \thegroup acting on a set \thespace
in any old way.  
While we might lose uniqueness and/or existence of vectorizations 
and parallelizations,
many of the arguments in this section would still go through.
However, using PHSes instead of general group actions
makes one thing particularly simple:
the proof of random self-reducibility for vectorization
and parallelization is identical to the usual arguments
for groups, which we sketched in~\S\ref{sec:DLP-CDHP}.
More precisely:
if an algorithm successfully solves vectorizations
in \((\thegroup,\thespace)\) with a probability of \(1/M\),
then we can solve \emph{any} vectorization in \((\thegroup,\thespace)\) 
with an expected \(M\) calls to the algorithm.
Given a target vectorization \((P,Q=\frakg\cdot P)\mapsto \frakg\),
we attempt to solve 
\((\fraka\cdot P,\frakb\cdot Q)\mapsto \fraka\frakb\frakg\)
for randomly chosen \(\fraka\) and \(\frakb\) in \thegroup;
we expect to land in the subset of inputs
where the algorithm succeeds within \(M\) attempts, 
and then recovering \frakg from 
\((\fraka,\frakb,\fraka\frakb\frakg)\) is trivial.
This means that the average- and worst-case difficulties for
vectorization are equivalent; a similar argument yields the same result
for parallelization.

Now, consider the relationship between vectorization and
parallelization.
If we can solve vectorizations
\((P,\frakg\cdot P) \mapsto \frakg\),
then we can can solve parallelizations
\((P,\fraka\cdot P,\frakb\cdot P) \mapsto \fraka\frakb\cdot P\),
so parallelization is notionally easier than vectorization.

As we have seen, the parallelization operation
\((P,\fraka\cdot P,\frakb\cdot P) \mapsto \fraka\frakb\cdot P\)
acts as the group law induced on \thespace by \thegroup
when elements are hidden by \(\varphi_P: \mathfrak{g}\mapsto\mathfrak{g}\cdot P\).
Given a parallelization oracle for
a PHS \((\thegroup,\thespace)\) 
with respect to one fixed base point \(P\)
(call this \emph{\(P\)-parallelization}),
we can view \(\thespace\) as an efficiently computable group,
and thus apply any black-box group algorithm to \(\thespace\).
Further, given an efficient \(P\)-parallelization algorithm,
the map \(\varphi_P\) becomes an efficiently computable group
homomorphism.
Even further, if we have a \(P\)-parallelization oracle 
for \emph{all} \(P\) in \thespace,
then the mapping \((\frakg,P)\mapsto\frakg\cdot P\)
becomes an efficiently computable bilinear pairing \(\thegroup\times\thespace\to\thespace\)
(viewing \thespace as a version of \thegroup hidden by one \(\varphi_O\)).

The efficient homomorphism \(\varphi_P:\thegroup\to\thespace\) 
implied by a \(P\)-parallelization oracle
is of course an isomorphism
(because \(\#\thespace = \#\thegroup\)),
but its inverse is not necessarily efficient\footnote{%
    The term \emph{one-way group action} is used for the HHS
    framework in~\cite{CSIDH} and~\cite{BonSchrot}.
    This hints at a more general setting,
    where actions are not necessarily simple or transitive.
}:
if it were,
then we could solve vectorizations \((P,\frakg\cdot P)\mapsto\frakg\)
because \(\frakg = \varphi_P^{-1}(\frakg\cdot P)\).
Conversely,
if we can vectorize with respect to \(P\),
then we can invert \(\varphi_P\):
the preimage \(\varphi_P^{-1}(T)\) of any \(T\) in \thespace
is the vectorization of \(T\) with respect to \(\varphi_P(1)\).
Parallelization therefore yields an efficient isomorphism in one direction,
while vectorization yields the inverse as well.

In the case where a group \(\thegroupa\) has prime order,
we can use a CDHP oracle
to view \(\thegroupa\) as a \emph{black-box field}
in the sense of Boneh and Lipton~\cite{Boneh--Lipton}.
Given a base point \(P\) in \(\thegroupa\),
each element \([a]P\) of \(\thegroupa\) 
is implicitly identified with its discrete logarithm \(a\).
The Diffie--Hellman operation \((P,[a]P,[b]P)\mapsto[ab]P\) 
becomes an implicit multiplication,
allowing us to view \(\thegroupa\) as a model of \(\FF_N\),
and thus to 
apply various subexponential and polynomial-time reductions
from the DLP to the CDHP in~\thegroupa
(as we noted in~\S\ref{sec:pre-quantum-DH}).
A parallelization oracle for \((\thegroup,\thespace)\),
however,
only allows us to view \(\thespace\) as a black-box group,
not a black-box field;
we therefore have no equivalent of the den Boer or Maurer reductions
in the HHS setting.

The separation between vectorization and parallelization therefore seems
more substantial than the somewhat thin and rubbery separation between
the DLP and CDHP.
However, we would still like to have some upper bounds for the hardness of these
problems. 
For vectorization, we can give some algorithms.

In the classical setting,
Couveignes noted that Shanks' baby-step giant-step (BSGS)
and Pollard's probabilistic algorithms
for DLPs in groups extend to vectorizations in PHSes.
Algorithm~\ref{alg:bsgs-X} is 
a BSGS analogue for a PHS \(\thespace\) under~\(\thegroup\).\footnote{%
    Algorithm~\ref{alg:bsgs-X} 
    becomes the usual BSGS 
    for DLPs in \(\thegroup = \subgrp{\frake}\)
    if we let \(\thespace = \thegroup\) 
    (with the group operation as the action), 
    let \(P = 1\), and let \(Q\) be the discrete log target.
}
Given \(P\) and \(Q = \frakg\cdot P\) in \(\thespace\)
and a generator \(\frake\)
for (a subgroup of) \(\thegroup\),
Algorithm~\ref{alg:bsgs-X}
computes an exponent \(x\) such that \(\frakg = \frake^x\), 
if it exists
(if \(\frakg\) lies outside the subgroup generated by \(\frake\),
then the algorithm will fail and return \(\bot\)).

\begin{algorithm}
    \caption{BSGS for a PHS \(\thespace\) under (a subgroup of) a group \(\thegroup\).}
    \label{alg:bsgs-X}
    \KwIn{Elements \(P\) and \(Q\) in \(\thespace\), and an element
    \(\frake\) in \(\thegroup\)}
    \KwOut{\(x\) such that \(Q = \frake^x\cdot P\), or \(\bot\)}
    \Fn{\BSGS{\(P,Q,\frake\)}}{
        \(\beta \gets \lceil{\sqrt{\#\subgrp{\frake}}}\rceil\)
        \tcp*{May be replaced with an estimate if \(\#\subgrp{\frake}\) not known}
        \(\mathcal{S} \gets \{\}\)
        \tcp*{Hash table: keys in \(\thespace\), values in \(\ZZ/N\ZZ\)}
        \(T \gets P\)
        \;
        \For{\(i\) in \([0,\ldots,\beta]\)}{
            \(\mathcal{S}[T] \gets i\)
            \;
            \(T \gets \frake\cdot T\)
            \;
        }
        \((T,\mathfrak{c}) \gets (Q,\frake^\beta)\)
        \;
        \For{\(j\) in \([0,\ldots,\beta]\)}{
            \If{\(T \in \mathcal{S}\)}{
                \(i \gets \mathcal{S}[T]\)
                \;
                \Return{\(i - j\beta\)}
                \tcp*{\(\frake^{j\beta}\cdot Q = \frake^i\cdot P\)}
            }
            \(T \gets \mathfrak{c}\cdot T\)
            \;
        }
        \Return{\(\bot\)}
    }
\end{algorithm}

Vectorization in a PHS \(\thespace\) under \(\thegroup\)
can always be solved classically in time and space \(\softO(\sqrt{\#\thegroup})\)
using Algorithm~\ref{alg:bsgs-X} and random self-reducibility,
provided a generator of a polynomial-index subgroup of \thegroup is known.
Algorithm~\ref{alg:bsgs-X} does more than what is required:
it returns not just the desired vectorization \(\frakg\),
but the discrete logarithm of \(\frakg\) with respect to \(\frake\).
This betrays the fact that Algorithm~\ref{alg:bsgs-X}
is just a black-box group algorithm
operating on a hidden version of \thegroup.

Pollard's algorithms also generalize easily to the HHS
setting, because we can compute the pseudorandom walks
using only translations, or ``shifts'', by group elements.
These translations in the group setting can be replaced by actions by
the same elements in the HHS setting.
The space complexity of vectorization can thus be reduced to 
as little as \(O(1)\) for the same time complexity.

Moving to the quantum setting:
despite its resemblance to the DLP,
vectorization cannot be solved with Shor's algorithm.
In fact, vectorization is an instance of 
the \emph{abelian hidden shift problem}~\cite{vDHI}:
given functions \(f\) and \(g\)
such that \(f(x\cdot s) = g(x)\) for all \(x\) and some ``shift''~\(s\), 
find \(s\).
The hidden shift instance corresponding to the
vectorization instance \((P,Q = \frakg\cdot P)\)
has \(f = \varphi_P:\thegroup\to\thespace\), 
\(g = \varphi_Q:\thegroup\to\thespace\),
and \(s = \frakg\).
Kuperberg reduces the abelian hidden shift problem
to an instance of the dihedral hidden subgroup problem,
which is then solved with a quantum algorithm
with a query complexity of \(L_N[1/2,c]\),
where \(c = \sqrt{2}\) according to~\cite{Childs--Jao--Soukharev}.
Kuperberg's original algorithm~\cite{Kup}
uses subexponential space;
Regev's simpler algorithm~\cite{regev04} uses polynomial quantum space;
and Kuperberg's most recent work~\cite{Kuperberg2013}
uses linear quantum space, but subexponential classical space.
More detailed perspectives on these algorithms
in the context of the isogeny class HHS
appear in~\cite{CSIDH,DFKS,BonSchrot}.

\section{
    The difficulty of exploiting subgroup structures
}
\label{sec:PH}

Moving back to the abstract:
when we think about DLPs, in black-box or in concrete groups,
we implicitly and systematically apply the Pohlig--Hellman--Silver algorithm
to reduce to the prime-order case.
It is interesting to note that for PHSes,
no such reduction is known:
it appears difficult to exploit 
the subgroup structure of \(\thegroup\)
when solving vectorization problems in \(\thespace\).\footnote{%
    It might seem odd that some black-box group algorithms 
    like BSGS and Pollard \(\rho\)
    adapt easily to PHSes,
    but not others like Pohlig--Hellman.
    But looking closer,
    BSGS and Pollard \(\rho\) 
    in groups
    only require translations,
    and not a full group law.
    We can therefore see BSGS and Pollard \(\rho\)
    not as black-box group algorithms,
    but rather as black-box PHS algorithms
    that are traditionally applied with \(\thespace = \thegroup\).
}

Algorithm~\ref{alg:PHS-G}
presents the Pohlig--Hellman--Silver algorithm for discrete logarithms 
in a group \thegroup whose order has known
factorization \(N = \prod_iN_i^{e_i}\).
Line~\ref{alg:PHS-G:DLP} applies a DLP-solving algorithm
(like
BSGS, Pollard \(\rho\),
or a specialized algorithm for a concrete group)
in the order-\(N_i\) subgroup of \thegroup.
If the factorization is complete and the \(N_i\) are prime,
then the global DLP is reduced to a polynomial number 
smaller prime-order sub-DLPs.

\begin{algorithm}
    \caption{Pohlig--Hellman--Silver 
        for a group \(\thegroup\)
        whose order has known (partial) factorization.
    }
    \label{alg:PHS-G}
    \KwIn{An element \(\frake\) of \thegroup,
        a \(\frakg\) in \(\subgrp{\frake}\),
        and
        \(((N_1,e_1),\ldots,(N_n,e_n))\)
        such that
        \(\#\thegroup = N = \prod_{i=1}^{n}N_i^{e_i}\),
        with the \(N_i\) pairwise coprime
        and the \(e_i>0\).
    }
    \KwOut{\(x\) such that \(\frakg = \frake^x\)}
    \Fn{\PohligHellman{\(\frake,\frakg,((N_1,e_1),\ldots,(N_n,e_n))\)}}{
        \For{\(1 \le i \le n\)}{
            \(\frake_i \gets \frake^{N/{N_i^{e_i}}}\)
            \label{alg:PHS-G:proj-i1}
            \;
            \(\frakg_i \gets \frakg^{N/{N_i^{e_i}}}\)
            \label{alg:PHS-G:proj-i2}
            \;
            \(x_i \gets 0\)
            \;
            \For{\(j\) in \((e_i-1,\ldots,0)\)}{
                \(\mathfrak{s} \gets {\frake_i}^{(N_i)^j}\)
                \label{alg:PHS-G:proj-j1}
                \tcp*{\(\mathfrak{s}\) is in the order-\(N_i\) subgroup}
                \(\mathfrak{t} \gets \mathfrak{s}^{-x_i}\cdot \frakg_i^{(N_i)^j}\)
                \label{alg:PHS-G:proj-j2}
                \tcp*{\(\mathfrak{t}\) is in the order-\(N_i\) subgroup}
                \(y \gets \log_{\mathfrak{s}}(\mathfrak{t})\)
                \label{alg:PHS-G:DLP}
                \tcp*{Use e.g.~baby-step giant-step}
                \(x_i \gets x_iN_i + y\)
                \;
            }
        }
        \(x \gets \CRT(\{(x_i,N_i^{e_i}) : 1 \le i \le n\})\)
        \;
        \Return{\(x\)}
        \;
    }
\end{algorithm}

The key steps involve producing the subgroup DLPs.
Lines~\ref{alg:PHS-G:proj-i1}
and~\ref{alg:PHS-G:proj-i2}
project the DLP instance \((\frakg,\frakh)\)
into the order-\(N_i^{e_i}\) subgroup of \(\subgrp{\mathfrak{g}}\).
Lines~\ref{alg:PHS-G:proj-j1}
and~\ref{alg:PHS-G:proj-j2}
then produce a DLP instance in the order-\(N_i\) subgroup.
This is always done by exponentiation by \(N/N_i^{e_i}\) and~\(N_i\);
indeed, this is the only way that the factors \(N_i\) are used in the algorithm.

In the PHS setting,
subgroup DLPs should be replaced with subgroup vectorizations.
Line~\ref{alg:PHS-G:DLP} could be replaced with a call to
Algorithm~\ref{alg:bsgs-X}, 
using \(\mathfrak{e}^{N_i^{e_i-1}}\) as the subgroup generator;
the problem is to produce a vectorization instance 
in a sub-PHS acted on by the corresponding subgroup.
We cannot naively concatenate ``\(\cdot P\)'' (or ``\(\cdot Q\)'') 
to most lines of the algorithm to turn group elements and operations
into PHS elements and operations:
Line~\ref{alg:PHS-G:proj-i2}, for example,
would require computing \(\frakg^{N/N_i^{e_i}}\cdot P\)
from \(Q = \frakg\cdot P\), \(P\), and \(\frakg\),
but this amounts to an iterated parallelization---%
and parallelization is supposed to be hard in an HHS.

A thorough investigation of the possibility and difficulty of exploiting
subgroup structures for vectorization and parallelization would require 
working with subgroup actions on quotient spaces;
we do not have room to discuss this here.
We note, however, that in some protocols
a limited number of exploitable parallelizations are provided
by the protocol itself, 
as in the group-based protocols subject to Cheon's attack~\cite{Cheon06},
and this should have some consequences for the security
of any HHS analogues of these protocols.

\section{
    A quick introduction to isogenies
}
\label{sec:isogeny-background}

This section provides enough background on
isogenies and endomorphisms of elliptic curves
to make sense of the HHS from Example~\ref{eg:isogeny-HHS}
before we describe cryptosystems based on it in~\S\ref{sec:isogeny-DH}.
We also want to fill in some background on supersingular curves
before we need it in~\S\ref{sec:SIDH}.
We assume a basic familiarity with the arithmetic of
elliptic curves;
readers familiar with isogenies and isogeny graphs
can safely skip this section.
As a mathematical reference, 
we suggest~\cite{silverman:elliptic} and~\cite{Kohel};
for greater detail focused on the cryptographic use case, see~\cite{DeFeo}.

We want to talk about relationships between elliptic curves
over a fixed finite field \(\FF_q\), where \(q\) is a power of some
prime \(p\).
We can assume that \(p \not= 2\) or \(3\), to simplify,
though the theory applies to those cases as well.
We will work with elliptic curves as short Weierstrass models
\(\EC: y^2 = x^3 + ax + b\), with \(a\) and \(b\) in~\(\FF_q\):
in practice we might compute using other curve models
(many isogeny-based cryptosystem implementations have preferred 
Montgomery arithmetic~\cite{Costello--Smith}),
but since we end up working with curves up to \(\FF_q\)-isomorphism,
and every curve is \(\FF_q\)-isomorphic to a short Weierstrass model,
we lose nothing in restricting to this simple and universal curve shape
in this article.
The \(m\)-torsion \(\EC[m]\) of \(\EC\)
is the subgroup of points \(P\) such that \([m]P = 0_\EC\).

A \emph{homomorphism} \(\phi: \EC \to \EC'\) is, 
by definition\footnote{%
    An elliptic curve is by definition a pair \((\EC,0_\EC)\),
    where \(\EC\) is a curve of genus 1 
    and \(0_\EC\) is a distinguished point on \(\EC\)
    (which becomes the identity element of the group of points;
    cf.~Example~\ref{eg:torsor});
    so it makes sense that a morphism 
    \((\EC,0_{\EC}) \to (\EC',0_{\EC'})\) 
    in the category of elliptic curves
    should be a mapping of algebraic curves \(\EC \to \EC'\)
    preserving the distinguished points,
    that is, mapping \(0_\EC\) onto \(0_{\EC'}\).
}, a rational map 
such that \(\phi(0_\EC) = 0_{\EC'}\).
Homomorphisms induce homomorphisms
on groups of points~\cite[\S{}III.4]{silverman:elliptic},
but not every homomorphism of groups of points
is induced by a homomorphism of curves.
An \(\FF_q\)-homomorphism is one that is defined over \(\FF_q\):
that is, the rational functions defining it as a rational map 
have coefficients in \(\FF_q\).
Every homomorphism here will be defined over \(\FF_q\),
unless explicitly stated otherwise.

\emph{Isogenies} are nonzero homomorphisms of elliptic curves.
If there is an isogeny from \(\EC\) to \(\EC'\),
then we say that \(\EC\) and \(\EC'\) are \emph{isogenous}.
We will see below that for each isogeny \(\EC \to \EC'\)
there is a dual isogeny \(\EC' \to \EC\),
so isogeny is an equivalence relation on elliptic curves.

\emph{Isomorphisms} are invertible homomorphisms.
The \(j\)-invariant of a curve \(\EC:y^2= x^3 + ax + b\)
is \(j(\EC) = 1728\cdot\frac{4a^3}{4a^3+27b^2}\);
two curves \(\EC\) and \(\EC'\) are \(\FFbar_q\)-isomorphic 
if and only if \(j(\EC) = j(\EC')\).
We need to work with the stronger notion of \(\FF_q\)-isomorphism,
where the \(j\)-invariant does not tell the whole story.
Curves that are \(\FFbar_q\)-isomorphic but not \(\FF_q\)-isomorphic
are called \emph{twists}.
The most important example is the \emph{quadratic twist},
which is isomorphic over \(\FF_{q^2}\) but not \(\FF_q\):
the quadratic twist of \(\EC: y^2 = x^3 + ax + b\)
is \(\EC': v^2 = u^3 + \mu^2au + \mu^3b\),
where \(\mu\) is any nonsquare in \(\FF_q\)
(the choice of nonsquare makes no difference up to \(\FF_q\)-isomorphism,
which is why we say \emph{the} rather than \emph{a} quadratic twist).
The isomorphism \(\tau: \EC \to \EC'\)
is defined by \((x,y) \mapsto (u,v) = (\mu x,\mu^{3/2}y)\);
this is clearly not defined over \(\FF_q\),
yet \(j(\EC) = j(\EC')\).
The quadratic twist of a curve \(\EC\) is its only twist,
up to \(\FF_q\)-isomorphism,
unless \(j(\EC) = 0\) or \(1728\) 
(in which case there may be four or two more twists, respectively).
Specifying an \(\FF_q\)-isomorphism class
therefore comes down to specifying a \(j\)-invariant and a choice of twist.

\emph{Endomorphisms} are homomorphisms from a curve to itself.
The endomorphisms of a given curve \(\EC\)
form a ring \(\End(\EC)\),
with the group law on \(\EC\) 
inducing addition of endomorphisms
and composition of endomorphisms corresponding with multiplication.
The structure of the set of isogenies from \(\EC\) to other curves
is deeply connected to the structure of \(\End(\EC)\),
and vice versa.

The scalar multiplication maps \([m]\) are endomorphisms,
so \(\End(\EC)\) always contains a copy of \(\ZZ\).
It also includes the \emph{Frobenius} endomorphism 
\(\pi: (x,y)\mapsto(x^q,y^q)\),
which satisfies the quadratic equation \(\chi(X) := X^2 - tX + q = 0\)
for some integer \(t\) in the \emph{Hasse interval}
\([-2\sqrt{q},2\sqrt{q}]\);
we call \(t\) the \emph{trace} of Frobenius (and of \(\EC\)).
Since points in \(\EC(\FF_q)\)
are precisely the points fixed by \(\pi\),
we have \(\#\EC(\FF_q) = \chi(1) = q + 1 - t\).
If \(\EC'\) is the quadratic twist of \(\EC\)
and we pull back the Frobenius on \(\EC'\)
to an endomorphism on \(\EC\) via the twisting isomorphism,
then the result is \(-\pi\),
so the trace of \(\EC'\) is the negative of the trace of \(\EC\).

Now consider the set of all elliptic curves over \(\FF_q\).
Tate's theorem tells us that two elliptic curves 
are \(\FF_q\)-isogenous if and only if they have the same trace
(and hence the same number of rational points).
This means that the set of all elliptic curves
is partitioned into \(\FF_q\)-isogeny classes,
indexed by the integers in the Hasse interval (via the trace).
Since the trace of a curve over \(\FF_q\)
is the negative of the trace of its quadratic twist,
and the quadratic twist is generally the only twist,
we can use the \(j\)-invariant to uniquely identify
elements of the isogeny class of trace \(t\not=0\) up to \(\FF_q\)-isomorphism,
even though \(j\) normally only classifies curves up to
\(\FFbar_q\)-isomorphism.
We can handle \(j = 0\) and \(1728\) as rare special cases,
but for the case \(t = 0\) we need to be more careful.

Now let us focus on a single \(\FF_q\)-isogeny class.
The isogeny class immediately breaks up into 
a union of \(\FF_q\)-isomorphism classes.
The modern way of looking at an \(\FF_q\)-isogeny class
is as a graph,
with \(\FF_q\)-isomorphism classes of curves for vertices,
and \(\FF_q\)-isomorphism classes of isogenies for edges
(isogenies \(\phi_1 : \EC_1 \to \EC_1'\) and \(\phi_2 : \EC_2\to\EC_2'\)
are isomorphic if there are isomorphisms \(\iota: \EC_1 \to \EC_2\)
and \(\EC_1'\to\EC_2'\) such that \(\phi_2\circ\iota = \iota'\circ\phi_1\)).

Tate's theorem is not constructive, so we generally don't know how to
get from one point to another in an isogeny graph.
The difficulty of computing a path representing an unknown isogeny
between given elliptic curves in the same isogeny class
is the source of most hard problems in isogeny-based cryptography.

To take a closer look at the structure of isogeny graphs
we need to classify isogenies, 
and to break them down to into fundamental components.
Our main tool for this is the \emph{degree}.
Since an isogeny \(\phi: \EC \to \EC'\) is defined by nonconstant rational maps,
it induces an extension 
\(\phi^\#: \FF_q(\EC')\hookrightarrow\FF_q(\EC)\) of function fields;
the degree \(\deg(\phi)\) of \(\phi\) is defined to be the degree of that extension.
(We extend the definition of degree to homomorphisms by 
defining the degree of zero maps to be 0.)
If \(\phi: \EC \to \EC'\) and \(\phi': \EC'\to\EC''\)
are isogenies, then \(\deg(\phi'\circ\phi) = \deg\phi\cdot\deg\phi'\).
Two examples are particularly important:
\(\deg[m] = m^2\), and \(\deg\pi = q\).
If \(\deg\phi = d\), then we say that \(\phi\) is a \emph{\(d\)-isogeny}.

Another important quality of isogenies is \emph{(in)separability},
which we define according to the (in)separability
of the corresponding function field extension.
For our purposes,
the purely inseparable isogenies are all iterated compositions of
\(p\)-powering \((x,y)\mapsto(x^p,y^p)\) (such as Frobenius);
these can be factored out of any other isogeny,
and then what remains is separable.

Suppose \(S\) is a finite subgroup of \(\EC(\FFbar_q)\).
Now \(S\) must include \(0\),
and it is also fixed by \([-1]\);
so \(S\) is determined precisely by the \(x\)-coordinates
of its nonzero elements.
We can therefore encode \(S\)
as a polynomial \(F_S(X) = \prod_P(X - x(P))\),
where the product runs over the nonzero points \(P\) of \(S\)
in such a way that \(P\) is included iff \(-P\) is not.
The subgroup \(S\) is defined over \(\FF_q\)
if and only if the polynomial \(F_S\) has coefficients in \(\FF_q\).

Being homomorphisms,
isogenies have kernels.
The kernel of an isogeny \(\phi: \EC \to \EC'\)
is always a finite subgroup of \(\EC\).
If \(\phi\) is separable,
then \(\#\ker\phi = \deg\phi\).
The points of \(\ker\phi\) are generally 
defined over an extension of \(\FF_q\),
but \(\ker\phi\) can be encoded as the \emph{kernel polynomial}
\(F_{\ker\phi}\), which is defined over \(\FF_q\).
Separable isogenies are defined by their kernels,
up to isomorphism.

Going in the other direction,
given a finite subgroup \(S\) of \(\EC\) defined over \(\FF_q\),
there exists a separable \emph{quotient isogeny}
\(\phi: \EC \to \EC/S\)
with \(\ker\phi = S\).
The isogeny and the curve \(\EC/S\)
are both defined up to \(\FF_q\)-isomorphism;
they can be computed using \emph{Vélu's formul\ae{}}~\cite{Velu71}.
(If \(S\) is encoded as the polynomial \(F_S\),
then we compute \(\phi_S\) using the symmetric version of Vélu's
formul\ae{} in~\cite[\S2.4]{Kohel}.)

Given an ideal \(\fraka\subset \End(\EC)\),
we can consider the subgroup 
\(\EC[\fraka] := \cap_{\psi\in\fraka}\ker\psi\).
This is the kernel of an isogeny
\(\phi: \EC \to \EC/\EC[\fraka]\);
the isogenies that arise in this way are central to
the key exchange of~\S\ref{sec:isogeny-DH}.
The degree of \(\phi\) is the norm of \(\fraka\) in \(\ZZ \subset \End(\EC)\).
If \(\fraka = (\psi)\) is principal,
then \(\phi\) is isomorphic to \(\psi\).

Given any \(d\)-isogeny \(\phi: \EC \to \EC'\),
we can compute the subgroup \(S = \phi(\EC[d]) \subset \EC'\),
and then the quotient \(\phi_S: \EC'\to\EC'/S\)
is a \(d\)-isogeny
such that \(\phi_S\circ\phi\) has kernel \(\EC[d]\);
hence, \(\phi_S\) is isomorphic to a \(d\)-isogeny
\(\dualof{\phi}: \EC' \to \EC\)
such that \(\dualof{\phi}\circ\phi = [d]\) on \(\EC\)
(and \(\phi\circ\dualof{\phi} = [d]\) on \(\EC'\)).
We call \(\dualof{\phi}\) the \emph{dual} of \(\phi\).
The upshot is that every \(d\)-isogeny \(\EC \to \EC'\)
has a corresponding \(d\)-isogeny \(\EC'\to\EC\).

Every isogeny can be factored into a composition of 
isogenies of prime degree, though there are two important caveats:
factorization is not unique, and generally a factorization may only
exist over some extension field.
For example, if \(\ell\not=p\) is prime,
then \(\EC(\FFbar_q)\cong(\ZZ/\ell\ZZ)^2\),
so there are \(\ell+1\) order-\(\ell\) subgroups \(S\subset \EC(\FFbar_q)[\ell]\),
each the kernel of a different isogeny \(\phi_S:\EC\to\EC/S\),
and then the dual isogeny gives us a factorization
\([\ell] = \dualof{\phi_S}\circ\phi_S\).
Each decomposition is only defined over the field of definition of the
associated subgroup~\(S\).

Just as we decompose isogenies into \(\ell\)-isogenies,
so consider the subgraphs formed by \(\ell\)-isogenies.
The structures of \(\ell\)-isogeny graphs 
depend strongly on the endomorphism rings
of curves in the isogeny class,
as we will see.

A curve \(\EC\) is \emph{supersingular} if \(p\) divides its trace 
(over \(\FF_p\), this implies the trace is \(0\)).
If \(\EC\) is not supersingular, then it is \emph{ordinary}.
The \(j\)-invariant of any supersingular curve is in \(\FF_p\) or \(\FF_{p^2}\),
so any supersingular curve is isomorphic to one defined over
\(\FF_p\) or \(\FF_{p^2}\).
There are roughly \(\lfloor{p/12}\rfloor\) supersingular
\(j\)-invariants in \(\FF_{p^2}\),
of which 
\(O(\sqrt{p})\) are in \(\FF_p\)
(more precisely, this number is the class number of \(\QQ(\sqrt{-p})\)).
Since supersingularity is defined by the trace,
either all of the curves in an isogeny class are supersingular,
or all of them are ordinary;
the two kinds of curves do not mix.

There are two possibilities for the general structure of the
endomorphism ring of an elliptic curve over a finite field:
\begin{description}
    \item[commutative]
        \(\End(\EC)\) is isomorphic to an order in a quadratic imaginary
        field; or
    \item[noncommutative]
        \(\End(\EC)\) is isomorphic to a maximal order in a quaternion algebra.
\end{description}
All ordinary curves have commutative endomorphism rings.
If a supersingular curve is defined over \(\FF_p\),
then its endomorphism ring is commutative\footnote{%
    If we consider endomorphisms defined over \(\FF_{p^2}\),
    then the ring is noncommutative.
};
if it is defined over \(\FF_{p^2}\),
then its endomorphism ring is noncommutative.

The commutative case is relatively simple:
each \(\End(\EC)\) is an order in \(K = \QQ(\pi)\) containing
the quadratic ring \(\ZZ[\pi]\).
The discriminant of \(\ZZ[\pi]\) is \(\Delta_\pi := t^2 - 4q = m^2\Delta_K\),
where \(\Delta_K\) is the fundamental discriminant of \(K\).
The algorithmic exploration of ordinary isogeny graphs
begins with Kohel's thesis~\cite[Chapter 4]{Kohel};
these graphs are now mainstream computational tools
in the arithmetic of elliptic curves and elliptic curve
cryptography~\cite{Galbraith,Fouquet--Morain,GHS,Jao--Miller--Venkatesan}.
The analogous theory for supersingular curves over \(\FF_p\),
whose endomorphism rings are commutative and thus behave like ordinary
curves,
was explored by Delfs and Galbraith~\cite{Delfs--Galbraith}.

If \(\phi: \EC \to \EC'\)
is an \(\ell\)-isogeny of endomorphism rings with commutative
endomorphism rings (with \(\ell\) prime), 
then there are three possibilities:
\(\End(\EC) \cong \End(\EC')\) (we say \(\phi\) is \emph{horizontal}),
\(\End(\EC) \subset \End(\EC')\) with index \(\ell\)
(we say \(\phi\) is \emph{ascending}),
or
\(\End(\EC) \supset \End(\EC')\) with index \(\ell\)
(we say \(\phi\) is \emph{descending}).
An \(\ell\)-isogeny can only be ascending or descending if \(\ell\)
divides the conductor \(m\) of \(\ZZ[\pi]\) in \(O_K\),
and an \(\ell\)-isogeny \(\phi: \EC \to \EC'\) can only be horizontal
if \(\End(\EC)\) and \(\End(\EC')\)
are locally isomorphic to the maximal order \(O_K\) of \(K\) at
\(\ell\):
that is, if
\(
    \End(\EC)\otimes\ZZ_\ell
    \cong
    \End(\EC')\otimes\ZZ_\ell
    \cong
    O_K\otimes\ZZ_\ell
\).
The \(\ell\)-isogenies of ordinary curves thus form ``volcano'' structures:
cycles of horizontal isogenies link the curves \(\EC\)
with \(\End(\EC) \cong O_K\),
and from each of these curves a regular tree grows downwards, 
with its leaves in the curves with \(\End(\EC) \cong \ZZ[\pi]\) 
(which is minimal).
The vertices with \(\End(\EC)\cong O_K\) have two horizontal \(\ell\)-isogenies
(or one or zero, if the cycle is degenerate),
and \(\ell-1\) descending isogenies;
each other vertex has one ascending and \(\ell\) descending isogenies,
except for the minimal vertices, which have no further descending isogenies.

From our perspective, what is most interesting about the commutative
case
is that the (isomorphism classes) of curves \(\EC\)
with \(\End(\EC) \cong O_K\)
form a PHS under the action of the class group of \(\Cl(O_K)\).
We met this PHS in Example~\ref{eg:isogeny-HHS}.

The noncommutative case is much more complicated,
and we will be much more brief here.
The algorithmic applications of the full supersingular isogeny graph
go back to Mestre and Oesterlé~\cite{Mestre},
and more detail appears in the second half of Kohel's
thesis~\cite[Chapter 7]{Kohel}.
In the non-commutative case,
the \(\ell\)-isogeny graph is \((\ell+1)\)-regular and connected,
and it is an expander graph.

\section{
    Commutative isogeny-based key exchange
}
\label{sec:isogeny-DH}

Recall the PHS
space from Example~\ref{eg:isogeny-HHS},
which Couveignes conjectured was an HHS:
fix a prime power \(q\) 
and an integer \(t\) with \(|t| \le 2\sqrt{q}\),
set \(\Delta := t^2 - 4q\),
and let \(\OO_K\) be the maximal order (the ring of integers) of the 
quadratic imaginary field \(K := \QQ(\sqrt{\Delta})\).
For this HHS,
\begin{itemize}
    \item
        the space \thespace is the set of isomorphism classes of
        elliptic curves over \(\FF_q\) of trace \(t\)
        whose endomorphism rings are isomorphic to \(\OO_K\);
    \item
        the group \thegroup is the ideal class group \(\Cl(\OO_K)\)
        of \(\OO_K\);
        and
    \item
        the action \((\fraka,\EC)\mapsto\fraka\cdot \EC\)
        is evaluated by computing the 
        isogeny \(\phi: \EC \to \EC/\EC[\fraka]\),
        and taking~\(\fraka\cdot \EC\) to be 
        the isomorphism class of~\(\EC/\EC[\fraka]\).
\end{itemize}
The cardinality \(N\) of \thegroup
is the class number of \(O_K\),
which is roughly \(\sqrt{\Delta_K}\),
where \(\Delta_K\) is the discriminant of \(K\)
(essentially the squarefree part of \(\Delta\)).
There is no point in not maximising \(N\)
with respect to \(q\),
so we should use \(t\)
such that \(\Delta\) is already a fundamental discriminant;
this forces all curves in the isogeny class to have
\(\ZZ[\pi] = \End(\EC) = O_K\),
and then \(N = \#\thegroup \sim \sqrt{q}\).

The vectorization and parallelization problems in this HHS
are expressed concretely in terms of computing paths in isogeny graphs.
The fastest known classical algorithms for vectorization and parallelization 
in this HHS are the generic square-root algorithms,
which run in time \(O(\sqrt[4]{q})\).
In the quantum world,
Childs, Jao, and Soukharev have defined
a subexponential quantum isogeny evaluation algorithm~\cite{Childs--Jao--Soukharev},
which in combination with Kuperberg's algorithm
gives a full subexponential quantum algorithm
for solving vectorization in this HHS.
This applies identically to the ordinary and commutative-supersingular cases.
Further analysis of this approach can be found in~\cite{BonSchrot}.

Couveignes defined a key exchange 
(essentially Algorithms~\ref{alg:HHS-KeyGen} and~\ref{alg:HHS-DH})
and an identification protocol in this HHS
in~\cite{Couveignes}.
These protocols were essentially unknown outside
the French community until
Rostovtsev and Stolbunov independently proposed a public key
encryption scheme based on the same HHS~\cite{Rostovtsev--Stolbunov}.
Stolbunov~\cite{Stol} then derived more protocols, including an interactive
key exchange scheme similar to Algorithms~\ref{alg:HHS-KeyGen} and~\ref{alg:HHS-DH}.
The only real difference between Couveignes' and Stolbunov's
cryptosystems
is in the sampling of private keys,
each representing one of the two approaches mentioned
in~\S\ref{sec:HHS-crypto}.
Couveignes uses a true uniform
random sampling over the whole of the keyspace,
then applies a lattice reduction-based algorithm
to produce an equivalent key whose action is efficiently computable.
Rostovtsev and Stolbunov sample keys from a subset of efficiently computable keys
whose distribution they conjecture to be close enough to the uniform
distribution on the entire group.

One particularly nice aspect of these schemes
is that key validation can be made simple and efficient
(see~\cite[\S5.4]{DFKS}),
so we can safely use the scheme for static key exchange.
Since the group action is simple and transitive,
every element of the space is a legitimate public key.
To validate a given \(x\) in \(\FF_q\) 
as a public key,
therefore,
it suffices to check that \(x\) is the \(j\)-invariant
of a curve with endomorphism ring \(O_K\).
We immediately construct a curve \(\EC\) with \(j\)-invariant \(x\),
and check that it has the right trace
(which amounts to checking that \(\EC(\FF_q)\) has the claimed
cardinality), switching to the quadratic twist if necessary.
This ensures that \(\End(\EC) \subseteq O_K\);
if \(t\) is chosen such that \(\ZZ[\pi] = O_K\),
then we are already done;
otherwise, we check \(\End(\EC) = O_K\)
using Kohel's algorithm~\cite{Kohel}.

Regardless of how the private key ideals are sampled,
by the time we want to use them in the group action
they are presented as factored ideals
\[
    \fraka = \prod_{i=1}^r\frakell_i^{e_i}
    \quad
    \text{with}
    \quad
    -B_i \le e_i \le B_i
    \,,
\]
where the \(\frakell_i\)
are distinguished prime ideals
whose corresponding \(\ell_i\)-isogenies can be evaluated very quickly.
If the cost of evaluating an isogeny associated with kernel
\(\EC[\frakell_i]\) (for a random \(\EC\) in the isogeny class)
is \(C_i\),
then the exponent bounds \(B_i\)
should be chosen in such a way 
that the cost of evaluation \(\sum_{i=1}^r B_iC_i\)
is minimised
while keeping the number of private keys
\(\prod_{i=1}^r(2B_i+1)\) big enough.

Suppose then that want to compute an \(\ell\)-isogeny
from an elliptic curve \(\EC/\FF_q\)
for some prime \(\ell \not= p\).
We consider two methods of computing \(\ell\)-isogenies here.
The classic approach
is based on modular polynomials.
An alternative approach based on Vélu's formul\ae{}
was originally proposed in~\cite{DFKS},
and subsequently used in~\cite{CSIDH}.
Both approaches are discussed in greater detail
in~\cite[\S3.2]{DFKS}.

First, consider the ``modular'' approach.
Recall that the \(\ell\)-th (classical\footnote{%
    We use classical modular polynomials here for simplicity,
    but alternative modular polynomials
    such as Atkin's, which have smaller degree,
    are better in practice.
    These degrees are still in \(O(\ell)\),
    so the asymptotic efficiency of this approach
    does not change.
}) 
modular polynomial \(\Phi_\ell(J_1,J_2)\)
is defined over \(\ZZ\),
is monic of degree \(\ell+1\)
in both \(J_1\) and \(J_2\),
and the roots of \(\Phi_\ell(j(\EC),X)\) in \(\FF_q\) 
are the \(j\)-invariants of the curves \(\ell\)-isogenous to
\(\EC\) over~\(\FF_q\).
In fact, the \(\FF_q\)-irreducible factors correspond 
to Galois orbits of \(\ell\)-isogenies
(or, equivalently, to Galois orbits of order-\(\ell\) subgroups of \(\EC\)).
To compute the \(\ell\)-isogenous curves up to isomorphism,
therefore,
we (pre)compute \(\Phi_\ell\) and reduce it modulo \(p\);
then we evaluate one variable at \(j(\EC)\),
and compute the roots in \(\FF_q\) of the resulting univariate polynomial.

There are \(\ell+1\) curves \(\ell\)-isogenous to \(\EC\)
over \(\FFbar_q\).
Of these isogenies, at most two can preserve the endomorphism ring.
If we choose \(\EC\)
such that \(\End(\EC)\) is the maximal order
and equal to \(\ZZ[\pi]\)
(or at least such that \(\ell\) does not divide the conductor of
\(\ZZ[\pi]\) in \(O_K\),
so \(\ZZ[\pi]\) is locally maximal)
then \(\Phi_\ell(j(\EC),X)\) will have only two roots,
corresponding precisely to these horizontal isogenies.
If this is the first step in a walk of \(\ell\)-isogenies
then we must determine which of the two is in the ``correct'' direction,
corresponding to the ideal;
we can do this by checking the eigenvalue of Frobenius on the kernel,
for example.
But if we have already started walking,
then there is no need to do this:
we know that the ``wrong'' isogeny is the dual of the preceding step,
so we just ignore the \(j\)-invariant of that curve.
The total cost of this approach is dominated by finding the roots
of \(\Phi_\ell(j(\EC),\FF_q)\),
which is \(O(\ell\log q)\) \(\FF_q\)-operations.

The alternative ``Vélu'' approach 
is to construct isogeny kernels explicitly,
and compute the corresponding isogeny steps
using Vélu's formul\ae{}~\cite{Velu71}.
The idea is simple:
suppose \(\EC(\FF_q)\) contains a point \(P_\ell\) of order \(\ell\);
we want to compute the isogeny \(\EC \to \EC/\subgrp{P_\ell}\)
with kernel generated by \(P_\ell\).
We can compute the kernel polynomial 
\(F(X) = \prod_{i=1}^{(\ell-1)/2}(X - [i]P_\ell)\)
in \(\softO(\ell)\) \(\FF_q\)-operations;
if \(P_\ell\) is defined over a small extension of \(\FF_q\),
say \(\FF_{q^k}\),
then the cost is \(\softO(k^2\ell)\) \(\FF_q\)-operations.
We can then apply Vélu's formul\ae{}
(as in~\cite{Velu71} or~\cite[\S2.4]{Kohel})
to compute an equation for \(\EC/\subgrp{P_\ell}\);
we do not need an expression for the isogeny itself.
The total cost is dominated by the cost of computing \(F\),
which is \(\softO(k^2\ell)\) \(\FF_q\)-operations.

The Vélu approach is much faster than the modular approach
when \(k^2\ll \log{q}\),
but it requires us to use isogeny classes of curves
with many small-order subgroups over very low-degree extensions.
Such curves are rare, 
and hard to find by exhaustive search:
constructing them presents similar challenges to the construction of
pairing-friendly curves (though here we want many small primes dividing
the order over a degree-\(k\) extension, rather than one big prime).
We might try to do better by using 
the CM method~\cite{Agashe--Lauter--Venkatesan,Sutherland2012},
which constructs elliptic curves
with a specified group order---but the CM method
only works when the discriminant \(\Delta_K\) of the maximal order
(and hence the class group of the maximal order)
is very small, because \(\#\Cl(O_K) \sim \sqrt{|-\Delta_K|}\).
This means that if we use the CM method to generate parameters,
then the private key space is far too small
for these cryptosystems to be secure.
In~\cite{DFKS}
curve parameters are selected by running an extensive search
to maximise the number of primes \(\ell\) with 
points in \(\EC[\ell](\FF_{q^k})\)
for smallish \(k\),
using the Vélu approach for these primes
and the modular approach for the others.
This gives a significant improvement over
the pure modular approach,
but the result is still far from truly practical.

CSIDH~\cite{CSIDH}
steps around this obstruction in an extremely neat way,
by switching to supersingular curves over \(\FF_p\).
Since their endomorphism rings are commutative quadratic orders,
these curves behave like ordinary curves,
and the Couveignes--Rostovtsev--Stolbunov protocol
carries over without modification.
However, the fact that these curves necessarily have order \(p+1\)
makes it extremely simple to control their group structure
and class group size
by appropriately choosing \(p\) from within the desired range.
This close control means that we can force all of the small primes to be
``Vélu'' with \(k = 1\), which results in a speedup 
that beats ordinary-curve constructions
like that of~\cite{DFKS} by orders of magnitude.
Key validation is also simpler for these curves.
We are unaware of any impact on security, negative or positive,
stemming from the use of supersingular curves
as opposed to ordinary curves;
so far, each attack described as targeting either 
CRS or CSIDH (e.g.~\cite{BonSchrot}) applies equally to the other.
CSIDH therefore represents a practical post-quantum Diffie--Hellman replacement,
though the development of efficient side-channel-aware implementations
of commutative isogeny protocols remains an open problem.

\section{
    Supersingular isogeny Diffie--Hellman
}
\label{sec:SIDH}

We conclude with a brief discussion of Jao and De Feo's
supersingular isogeny Diffie--Hellman, known as SIDH~\cite{SIDH1,SIDH2}.
On the surface, SIDH resembles the commutative isogeny key exchange
of~\S\ref{sec:isogeny-DH}: 
Alice and Bob each compute a sequence of isogenies
to arrive at their public keys, and later the shared secret.
However, the differences are striking.

The most fundamental difference is that 
the endomorphism rings in SIDH are noncommutative,
so the algebraic objects acting on the isogeny class are not abelian
groups:
SIDH falls squarely outside the HHS framework.
In particular,
Alice and Bob's isogeny walks
do not automatically commute;
some extra data must be passed around
to correctly orient their walks for the second phase of the key exchange.

The second crucial difference with commutative isogeny key exchange
is that the underlying \(\ell\)-isogeny graphs
are no longer cycles;
rather, each is \((\ell+1)\)-regular, connected,
and an expander graph.
Since there are \(\Theta({p})\) vertices,
computing random sequences of \(O(\log p)\) \(\ell\)-isogenies
from a given base curve
takes us to a distribution of curves that we expect to be close 
to a uniform random distribution on the isogeny class.

To define the protocol,
we fix distinct primes \(\ell_A\) and \(\ell_B\)
(these will be very small, typically \(2\) and~\(3\)),
and exponents \(n_A\) and \(n_B\), respectively,
and let \(p\) be a prime such that
\(p = c\cdot \ell_A^{n_A}\cdot \ell_B^{n_B} \pm 1\)
for some very small \(c\).
We want to choose \(\ell_A\) and \(\ell_B\)
such that \(\ell_A^{n_A}\) and \(\ell_B^{n_B}\)
are roughly the same size;
ideally, \(\ell_A^{n_A} \sim \ell_B^{n_B} \sim \sqrt{p}\).

Now consider the supersingular isogeny class
over \(\FF_{p^2}\):
every curve \(\EC\) in it has 
\(\EC[\ell_A^{n_A}]\cong(\ZZ/\ell_A^{n_A}\ZZ)^2\)
and
\(\EC[\ell_B^{n_B}]\cong(\ZZ/\ell_B^{n_B}\ZZ)^2\).
Fix a base curve \(\EC_0\) in the isogeny class,
along with bases \((P_A,Q_A)\) of \(\EC_0[\ell_A^{n_A}]\)
and \((P_B,Q_B)\) of \(\EC_0[\ell_B^{n_B}]\).

First, key generation.
Alice samples a random \(a\) in \(\ZZ/\ell_A^{n_A}\ZZ\) as her private key;
the point \(P_A + [a]Q_A\) has exact order \(\ell_A^{n_A}\),
and generates the kernel of an \(\ell_A^{n_A}\)-isogeny 
\(
    \phi_A:\EC_0 \to \EC_A \cong \EC_0/\subgrp{P_A+[a]Q_A}
\),
which she computes as a series of \(\ell_A\)-isogenies.
Her public key is \((\EC_A,\phi_A(P_B),\phi_A(Q_B))\).
Bob samples a private key \(b\) in \(\ZZ/\ell_B^{n_B}\ZZ\)
and computes 
the \(\ell_B^{n_B}\)-isogeny
\(
    \phi_B:\EC_0 \to \EC_B \cong \EC_0/\subgrp{P_B+[b]Q_B}
\)
as a series of \(\ell_B\)-isogenies;
his public key is \((\EC_B,\phi_B(P_A),\phi_B(Q_A))\).
There is plenty of redundant information in these public keys,
and they can be compressed following the suggestions
in~\cite{CJLNRU}.

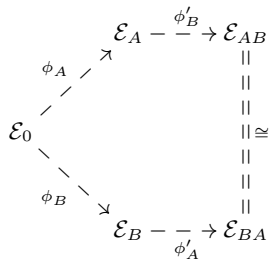
\begin{figure}
    \[
        \xymatrix{
            & 
            \EC_A
            \ar@{-->}[r]^{\phi_B'}
            &
            \EC_{AB}
            \ar@{==}[dd]^{\cong}
            \\
            \EC_0
            \ar@{-->}[ru]^{\phi_A} 
            \ar@{-->}[rd]_{\phi_B} 
            &
            &
            \\
            &
            \EC_B
            \ar@{-->}[r]_{\phi_A'}
            & 
            \EC_{BA}
        }
    \]
    \caption{Supersingular Isogeny Diffie--Hellman.}
    \label{fig:SIDH}
\end{figure}

To complete the key exchange,
Alice computes the \(\ell_A^{n_A}\)-isogeny 
\(\phi_A': \EC_{B} \to \EC_{BA} = \EC_B/\subgrp{\phi_B(P_A)+[a]\phi_B(Q_A)}\),
and 
Bob computes the \(\ell_B^{n_B}\)-isogeny
\(\phi_B': \EC_{A} \to \EC_{AB} = \EC_A/\subgrp{\phi_A(P_B)+[b]\phi_A(Q_B)}\).
The shared secret is the \(j\)-invariant
\(j(\EC_{AB}) = j(\EC_{BA})\) in \(\FF_{p^2}\);
the curves \(\EC_{AB}\) and \(\EC_{BA}\) have the same \(j\)-invariant
because both are isomorphic to
\(\EC_0/\subgrp{P_A + [a]Q_A + P_B + [b]Q_B}\).
The relationships between these curves is illustrated 
in Figure~\ref{fig:SIDH}.

Jao, De Feo, and Plût specified efficient algorithms for SIDH
in~\cite{SIDH2}.
The first competitive public implementation 
was due to Costello, Longa, and Naehrig~\cite{SIDH}.
A lot of effort has since been put into improving 
the algorithmic and space efficiency of
SIDH~\cite{CJLNRU,CostelloH17,FOR17}, 
and optimizing arithmetic in its specialized finite fields~\cite{BF1,BF2}.
One particlarly nice feature of SIDH in comparison to
commutative isogeny DH is that the isogenies in SIDH
all have degree either \(\ell_A\) or \(\ell_B\);
both are fixed, and typically tiny,
so computing the individual isogeny steps
is much faster in the supersingular protocol,
and requires much less code and precomputation.

Recovering the private key from an SIDH public key---%
a noncommutative analogue of vectorization---%
amounts to computing an isogeny between the base curve \(\EC_0\)
and the target public key curve \(\EC_A\).
We can do this using an algorithm
due to Delfs and Galbraith~\cite{Delfs--Galbraith},
inspired by an algorithm for the ordinary case
due to Galbraith, Hess, and Smart~\cite{GHS}.
The algorithm walks randomly through the supersingular isogeny graph
from the starting and ending curves,
until curves defined over \(\FF_p\) are detected---and then finding a
path between those two curves, to complete the desired isogeny,
is analogous to breaking a CSIDH key
(though with completely different security parameters).
Alternatively, 
Adj, Cervantes--Vázquez, Chi--Dominguez, Menezes, and
Rodríguez--Henríquez have given a useful analysis of
the van Oorschot--Wiener algorithm applied to this problem~\cite{ACCMR}.
The asymptotic cost of either approach is in \(O(\sqrt[4]{p})\)
\(\FF_{p^2}\)-operations
if \(\ell_A^{n_A} \sim \ell_B^{n_B} \sim \sqrt{p}\).
In the quantum setting,
we can apply Tani's claw-finding algorithm~\cite{Tani09}
to find a curve in the intersection of the sets of curves
\(\ell_A^{n_A/2}\)-isogenous to \(\EC_0\) and \(\EC_A\)
with a query complexity of \(O(\ell_A^{n_A/3})\)
(or we can attack Bob's public key in \(O(\ell_B^{n_B/3})\)),
which is \(O(p^{1/6})\)
when \(\ell_A^{n_A} \sim \ell_B^{n_B} \sim \sqrt{p}\).
The fact that the subexponential Childs--Jao--Soukharev algorithm
does not apply in the noncommutative case was one of the 
motivations for developing SIDH.

But SIDH keys do not simply present the target curve of an unknown
isogeny: they also present images of distinguished torsion bases,
which may help cryptanalysis~\cite{Petit17}.
The precise nature of the cryptographic problems underlying SIDH
is quite complicated, but Urbanik and Jao's survey of these problems
provides useful analysis~\cite{Urbanik--Jao},
while Eisentraeger, Hallgren, Lauter, Morrison, and Petit
go further into the connections with the endomorphism ring~\cite{EHLMP}.

Finding an isogeny between two supersingular curves over \(\FF_{p^2}\) 
is equivalent to determining their endomorphism rings,
under reasonable heuristics~\cite{EHLMP,Kohel,KLPT}.
This makes an interesting contrast with the commutative case,
where the endomorphism rings are presumed known,
and in any case can be computed using Kohel's
algorithm~\cite{Kohel}.  
As we have seen, 
determining the endomorphism ring 
is an important step in public key validation in commutative isogeny key exchange.

Key validation is especially problematic for SIDH.
Suppose we have an algorithm which,
given a prime \(\ell\), a positive integer \(n\),
and a curve \(\EC\),
efficiently decides whether \(\EC\) is \(\ell^n\)-isogenous to \(\EC_0\).
Such an algorithm would allow us to verify
whether Alice or Bob's public key was honestly generated 
(by calling the algorithm on
\((\ell_A,n_A,\EC_A)\)
or
\((\ell_B,n_B,\EC_B)\),
respectively).
However,
as we see in~\cite[\S6.2]{Galbraith--Vercauteren}
and~\cite{Thormarker},
this algorithm can also be used to efficiently recover secret keys
from public keys.
Indeed, 
take Alice's public curve \(\EC_A\);
there are \(\ell_A+1\) curves \(\ell_A\)-isogenous to it.
Computing each of these isogenies \(\phi: \EC_A\to \EC_A'\),
we call the algorithm on \((\ell_A,n_A-1,\EC_0,\EC_A')\);
if it returns true, then \(\phi\) is the last \(\ell_A\)-isogeny in
Alice's secret key.
Iterating this procedure reveals the entire key.

Problematic key validation 
makes defining a CCA-secure SIDH-based KEM
more complicated than the equivalent in the commutative case.
SIKE~\cite{SIKE},
which is the only isogeny-based candidate KEM in the NIST process,
handles this by applying 
the Hofheinz--Hövelmanns--Kiltz
a variant of the Fujisaki--Okamoto transform~\cite{HHK,Fujisaki--Okamoto}
to SIDH;
this entails a nontrivial performance hit.

On a formal level,
there are some profound differences between SIDH
and classical Diffie--Hellman.
The most obvious is the lack of symmetry in SIDH
between Alice and Bob,
whose roles are no longer interchangeable.
This is reflected by their distinct and incompatible key spaces, 
which are in turn distinct from the shared secret space 
and the space the base curve lives in.
Alice's private key encodes a sequence of \(\ell_A\)-isogenies 
of length \(n_A\),
while Bob's encodes a sequence of \(\ell_B\)-isogenies of length \(n_B\).
Alice's public key belongs to the space of (isomorphism classes of)
elliptic curves equipped with a distinguished \(\ell_B^{n_B}\)-torsion
basis,
while Bob's is equipped with an \(\ell_A^{n_A}\)-torsion
basis instead.
The base curve \(\EC_0\) is drawn from yet another space:
it is equipped with an \(\ell_A^{n_A}\ell_B^{n_B}\)-torsion basis.

This asymmetry might seem like a curious but minor inconvenience:
the participants just need to decide who is Alice and who is
Bob before each key exchange.
More importantly, though,
this asymmetry is incompatible with most of the theoretical machinery
that we use to reason about Diffie--Hellman and its hardness.
We have already seen how 
group Diffie--Hellman oracles create black-box field structures
on prime-order groups,
while HHS Diffie--Hellman oracles create a black-box group structures.
In SIDH, however,
a Diffie--Hellman oracle defines no binary operation on any set,
let alone an interesting algebraic structure.
This plurality of spaces makes it hard to adapt
hidden-number-problem-style arguments~\cite{Boneh--Venkatesan,Akavia09}
for hardcore bits to the SIDH context in a natural way,
though a valiant effort has been made by Galbraith, Petit, Shani, and Ti~\cite{GalbraithPST16}.

\begin{remark}
    At first glance, the fact that SIKE is the only isogeny-based KEM
    submitted to the NIST post-quantum process, competing with 58 others
    mostly based on codes, lattices, and polynomial systems, might suggest
    that it is a strange outlier.
    However, this uniqueness is not so much an indicator of lack of support,
    so much as a sign of rare convergence and consensus in the
    elliptic-curve cryptography community---convergence that did not occur
    to the same extent in the communities working on other post-quantum paradigms.
    The fact that there was only one isogeny-based submission
    reflects the general agreement
    that this was the right way to do isogeny-based key agreement 
    at that point in time.
    The more flexible CSIDH scheme was not developed until later, when the NIST process
    was already underway, and so it was not part of the conversation.
\end{remark}

\bibliographystyle{abbrv}
\bibliography{refs}

\end{document}